\documentclass[showpacs,preprintnumbers,amsmath,aps,nofootinbib,amssymb,superscriptaddress]{revtex4}
\usepackage{graphicx}
\usepackage{dcolumn}
\usepackage[english]{babel}
\usepackage{bm}
\usepackage[font=scriptsize]{caption}
\usepackage{ragged2e}
\usepackage[font=small,labelfont=bf,justification=raggedright]{caption}
\usepackage[utf8]{inputenc}
\usepackage[tikz]{bclogo}
\usepackage[framemethod=tikz]{mdframed}
\usepackage{amsmath}

\usepackage{mathtools}
\usepackage{xcolor}
\usepackage{lscape}
\usepackage{rotating}
\colorlet{BLUE}{blue} \colorlet{RED}{red}
\newsavebox{\measurebox}

\begin{document}

\title[]{Classical and quantum exact solutions for a FRW  in chiral like cosmology}
\author{J. Socorro}
\email{socorro@fisica.ugto.mx}
 \affiliation{Departamento de
F\'{\i}sica, DCeI, Universidad de Guanajuato-Campus Le\'on, C.P.
37150, Le\'on, Guanajuato, M\'exico}

\author{S. P\'erez-Pay\'an}
\email{saperezp@ipn.mx} \affiliation{Unidad Profesional
Interdisciplinaria de Ingenier\'ia,
Campus Guana\-jua\-to del Instituto Polit\'ecnico Nacional.\\
Av. Mineral de Valenciana \#200, Col. Fraccionamiento Industrial
Puerto Interior, C.P. 36275, Silao de la Victoria, Guana\-jua\.to,
M\'exico.}

\author{Rafael Hern\'andez-Jim\'enez}
\email{rafaelhernandezjmz@gmail.com}
 \affiliation{Departamento de
F\'isica, Centro Universitario de Ciencias
Exactas e Ingenier\'ia, Universidad de Guadalajara.\\
Av. Revoluci\'on 1500, Colonia Ol\'impica C.P. 44430, Guadalajara,
Jalisco, M\'exico.}

\author{Abraham Espinoza-Garc\'ia}
\email{aespinoza@ipn.mx} \affiliation{Unidad Profesional
Interdisciplinaria de Ingenier\'ia,
Campus Guana\-jua\-to del Instituto Polit\'ecnico Nacional.\\
Av. Mineral de Valenciana \#200, Col. Fraccionamiento Industrial
Puerto Interior, C.P. 36275, Silao de la Victoria, Guana\-jua\.to,
M\'exico.}

\author{Luis Rey D\'iaz-Barr\'on}
\email{lrdiaz@ipn.mx} \affiliation{Unidad Profesional
Interdisciplinaria de Ingenier\'ia,
Campus Guana\-jua\-to del Instituto Polit\'ecnico Nacional.\\
Av. Mineral de Valenciana \#200, Col. Fraccionamiento Industrial
Puerto Interior, C.P. 36275, Silao de la Victoria, Guana\-jua\.to,
M\'exico.}

\begin{abstract}
In this work, first, we study a flat Friedmann-Robertson-Walker
Universe  with two scalar fields but only one potential term, which
can be thought as a simple quintessence plus a K-essence model.
Employing the Hamiltonian formalism we are able to obtain the
classical and quantum solutions. The second model studied, is also a
flat Friedmann-Robertson-Walker Universe with two scalar fields,
with the difference that the two potentials are considered as well
as the standard kinetic energy and the mixed term (chiral field
approach). Regarding this second model, it is shown that setting to
zero the coefficient accompanying the mixed momenta term, two
possible cases can be studied: a quintom like case ($m^{12}_{+}$)
and a quintessence like case ($m^{12}_{-}$). For both scenarios
classical and quantum solutions are presented.
\end{abstract}

\maketitle

\section{Introduction}
One of the main goals of modern cosmology is to be able to adequately describe the early Universe.
In this sense, the first proposal for such a description was the Big-Bang theory; unfortunately this theory suffered
from two problems: that of flatness and that of the horizon. At the beginning of the 80's of the last century,
the idea of inflation was introduced \cite{guth1981, linde1982, turner1981,
starobinsky1980}, healing the problems that the Big-Bang theory had. Boldly speaking, the inflation process is a
period of exponential growth in our Universe. During this process, in addition to solving the problems already
mentioned, the inflation mechanism also explains the homogeneity and isotropy currently observed in the Universe.
Another important aspect of inflation is that the fluctuations generated during this period give rise to a primordial
 spectrum of density perturbations \cite{Starobinsky:1979ty,Mukhanov:1981xt,kodama,
bassett} which is nearly scale invariant, adiabatic and Gaussian,
and is in agreement with cosmological observations
\cite{Planck,OSR}.

On the other hand, scalar fields have been extensively used in the
past three decades as the possible major matter components for the
evolution of the Universe. They can describe various phenomena of
our Universe such as the inflationary era, the late time
acceleration, the dark matter component of the Universe and the
unification of early inflation to late acceleration, to mention a
few \cite{linde1982, Linde1983, Barrow1993_1, Barrow1993_2,
Peebles1987, Tsujikawa2013, Liddle1998, Sahni1999, Matos2000,
Urena-Lopez2016, Peebles1998, deHaro:2016_1, deHaro:2016_2,
Elizalde:2004mq}  (in the sense that what determines the
inflationary model is the form of the potential). From a
phenomenological point of view, the most successful models have been
those that have incorporated quintessence scalar fields and
slow-roll inflation \cite{OSR, Liddle1998, barrow, ferreira,
copeland1, copeland2, copeland3, andrew2007, gomez, capone,
kolb,Vagnozzi}, chiral cosmology connected to $f(R)$ theories or
nonlinear sigma model
\cite{chervon1995,Chervon2013,Kaiser2014,Fomin2017,Chervon2019,Paliathanasis2019,Paliathanasis2020a,Paliathanasis2020b,Bamba:2012cp,Dimakis:2020tzc,Dimakis:2019qfs,Paliathanasis:2014yfa}.
In general, scalar field theories use a single scalar field with its
respective potential; however, in recent years there have been
proposals where scalar field theories consist of multiple scalar
fields and which have yielded quite interesting results within the
context of the evolution of the Universe. For example, it has been
found that a system consisting of two scalar fields can describe the
crossing of the cosmological constant boundary ``$-1$'' (known as
quintom models \cite{Cai2009, Setare2008, Lazkoz2007, Leon2018}),
which for a single scalar field model is impossible since the single
scalar field models can only describe either the quintessence or the
phantom regime. Furthermore, these multi-field models can also
explain the early inflationary era of the Universe, known as hybrid
inflation \cite{chimento,lindle, cope, kim,omar-epjp2017} and gives
a different graceful exit in comparison with the standard
inflationary paradigm \cite{Wands2008, Bond2006, Inomata2017}. In
addition, the dynamical possibilities in multi-field inflationary
scenarios are considerably richer than in single-field models, such
as in the primordial inflation perturbations analysis
\cite{Yokoyama:2007dw, Chiba:2008rp} or the assisted inflation as
discussed in \cite{andrew1998a, Copeland:1999cs}.

In these models, as in the single field ones, the potential
associated to the scalar fields plays a very important role. In
several cases the potential that is employed is a simple exponential
product of the scalar fields or a series of linear sum exponentials
\cite{adrianov}. For example, in \cite{soco2, sor} a potential of
the form $\rm V(\phi,\sigma)=V_0\, e^{-\lambda_1 \phi -\lambda_2
\sigma}$ was employed. This potential was found under the connection
between the time derivatives of the momenta, namely,
$\dot\Pi_\phi\propto\dot\Pi_\sigma$, provided that $\partial
V/\partial\phi=\alpha\partial V/\partial\sigma$, (this type of
potentials have also been found under other considerations
\cite{OSR,omar-epjp2017}). Following this line of thought, first, we
shall consider the case when both potentials are proportional
between them, leaving the theory with only one potential but both
kinetic terms, and show that the exact analytical solution is
obtained employing the mathematical tools of Hamilton's formalism.
Then, the other scenario that we will investigate is when both
potentials come into play, that is, $\rm V(\phi,\sigma)=V_1\,
e^{-\lambda_1 \phi}+V_2 e^{ -\lambda_2\sigma}$ (this class of
potential in the scalar fields yields the so called chiral cosmology
(nonlinear sigma model) \cite{chervon1995}); in this setting,
because the two potentials are considered a mixed kinetic term has
to be introduced with a coupling parameter which allows us to obtain
a constraint in order to separate the equations and be able to solve
the problem analytically. The advantage of this scenario is that we
can reproduce the quintessence or quintom cosmologies choosing
appropriately the sign of this mixed kinetic term. These types of
scalar fields models have been introduced in the literature in order
to provide an alternative description to primordial inflation
\cite{Paliathanasis2019,Paliathanasis2020a,chimento,lindle,cope,kim,giacomini,benisty,faraoni,sivanesan,gorini,show,leon1}.

For the quantum cosmological cases we implement a basic formulation
by means of the Wheeler-DeWitt (WDW) equation. In order of being
able to obtain solutions to the WDW equation, several approaches
have been studied, such is the case of \cite{Gibbons}, where a
debate of what a typical wave function of the Universe is presented.
In \cite{Zhi}, a review on quantum cosmology where the problem of
how the Universe emerged from Big Bang singularity can no longer be
neglected in the GUT epoch is discussed. Moreover, the best
candidates for quantum solutions are those that have a damping
behavior with respect to the scale factor, since only such wave
functions allow for good solutions when using a
Wentzel-Kramers-Brillouin (WKB) approximation for any scenario in
the evolution of our Universe \cite{HH,H}. Furthermore, in the
context of a single scalar field a family of scalar potentials is
obtained in the Bohmian formalism \cite{omar-epjp2017,wssa}, or
supersymmetric quantum cosmology \cite{SPN,socorro3,socorro4}, where
among others a general potential of the form $\rm V(\phi)=V_0
e^{-\lambda \phi}$ is examined.

This work is arranged as follows. In section \ref{model} we will
introduce the model with two scalar fields, where both kinetic terms
are taken into account but only one term of the scalar potential is
present. From the corresponding Einstein-Klein-Gordon (EKG)
equations and the Hamiltonian density, we are able to obtain three
different solutions for the model. Also, into this same model we do
the analysis for different types of standard matter (stiff matter
and radiation) obtaining their corresponding solutions and present
the dust scenario for arbitrary scalar potential, whose particular
form obtained in the solution, plays an important role in the
behavior of the volume function. In section \ref{model-two}, we
analyze a two scalar field cosmological model but where both scalar
potentials come in to play. As will be shown, within this model we
can distinguish between two scenarios: a quintom like case and a
quintessence like case. For both scenarios we will present the
Hamilton equations and their corresponding solutions. Next, in
section \ref{qsolutions}, we consider the quantum versions of the
previous cosmological models calculating their corresponding WDW
equations and its solutions. Finally, section \ref{conclusions} we
give our final remarks.

\section{First Model \label{model}}
Let's start by introducing the multi-field Lagrangian density that
we will be working with, and can be thought of as standard
quintessence plus a simple K-essence model, including the standard
matter, thus we have

\begin{equation}
\rm {\cal L}=\sqrt{-g} \left( R-\frac{1}{2}g^{\mu\nu} \nabla_\mu
\phi_1 \nabla_\nu \phi_1 -\frac{1}{2}g^{\mu\nu} \nabla_\mu \phi_2
\nabla_\nu \phi_2 + V(\phi_1)\right) +\sqrt{-g} {\cal L}_{matter}
\,, \label{first-lagra}
\end{equation}
where $\rm R$ is the Ricci scalar, $\rm V(\phi_1)$ is the
corresponding scalar field potential  and ${\cal L}_{matter}$
corresponds at the contribution of ordinary matter for barotropic
perfect fluid, $\rm P=\gamma \rho$, with $\rho$ the energy density,
$\rm P$ is the pressure of the fluid in the co-moving frame,
$\gamma$ is the barotropic constant,  the equation of state to
scalar field is $\rm P_\phi=\omega_\phi \rho_\phi$, with the
pressure and energy density of the scalar field are defined in the
standard way in the gauge N=1, $\rm 16\pi G P_\phi=\frac{1}{2}{\dot
\phi}^2-V(\phi)$ and $\rm 16\pi G \rho_\phi=\frac{1}{2}{\dot
\phi}^2+V(\phi)$, however in the gauge $\rm N\not=1$, this is
defined with the modification in $\rm V(\phi)\to N^2 V(\phi)$ and
$\rho \to N^2 \rho$, and the reduced Planck mass $M_{P}^{2}=1/8\pi
G=1$.  Before we continue, it is important to mention that in
principle the potential in (\ref{first-lagra}) should be of the form
$V(\phi_1,\phi_2)$, but as has already been expressed, this form of
the Lagrangian density is obtained when one considers that the
potentials are proportional between them, leaving the theory with
only one dynamical potential, and two kinetic terms. To obtain the
corresponding EKG field equations we must perform the variations of
Eq.(\ref{first-lagra}) with respect to the metric and the scalar
fields, also we include the conservation law of the energy-momentum
tensor of a perfect fluid of ordinary matter, giving
\begin{align}
\rm G_{\alpha \beta}=\rm -T_{\alpha \beta}-\frac{1}{2}
\left(\nabla_\alpha \phi_1 \nabla_\beta \phi_1 -\frac{1}{2}g_{\alpha
\beta} g^{\mu \nu}
\nabla_\mu \phi_1 \nabla_\nu \phi_1 \right) +\frac{1}{2}g_{\alpha \beta} \, V(\phi_1)\nonumber\\
\rm -\frac{1}{2} \left(\nabla_\alpha \phi_2 \nabla_\beta \phi_2
-\frac{1}{2}g_{\alpha \beta} g^{\mu \nu}
\nabla_\mu \phi_2 \nabla_\nu \phi_2 \right), \label{munu}\\
\rm \Box \phi_1 -\frac{\partial V}{\partial \phi_1} =\rm
g^{\mu\nu} {\phi_1}_{,\mu\nu} - g^{\alpha \beta} \Gamma^\nu_{\alpha
\beta} \nabla_\nu \phi_1 - \frac{\partial V}{\partial \phi_1}=\rm 0
\,,\label{ekg-phi-1}\\
\rm g^{\mu\nu} {\phi_2}_{,\mu\nu} - g^{\alpha \beta}
\Gamma^\nu_{\alpha \beta} \nabla_\nu \phi_2=0\label{ekg-phi2},\\
\rm \nabla_\nu T^{\mu \nu}=0, \qquad T_{\mu \nu}=(\rho +P)u_\mu
u_\nu +g_{\mu \nu} P \label{tmunu}.
\end{align}

As we are considering a flat FRW Universe, the line element to be
used in this work is
\begin{equation}
\rm ds^2=-N(t)^2 dt^2 +e^{2\Omega(t)} \left[dr^2
+r^2(d\theta^2+sin^2\theta d\phi^2) \right], \label{frw1}
\end{equation}
where $\rm N$ is the lapse function, $\rm A(t)=e^{\Omega(t)}$ is the
scale factor in the Misner parametrization and  $\rm \Omega$ a
scalar function with interval from $-\infty$ to $\infty$. The
Einstein-Klein-Gordon field equations and conservation law of the
energy-momentum tensor of a perfect fluid of ordinary matter, are
\begin{align}
\rm \frac{3\dot{\Omega}^{2}}{N^2}-\frac{\dot{\phi_1}^2}{4N^2}-\frac{\dot{\phi_2}^2}{4N^2}-\frac{1}{2}V(\phi_1)-\rho=0,\label{ekg0}\\
\rm\frac{2\ddot{\Omega}}{N^2}+\frac{3\dot{\Omega}^2}{N^2}-\frac{2\dot{\Omega}\dot{N}}{N^3}+\frac{\dot{\phi_1}^2}{4N^2}+\frac{\dot{\phi_2}^2}{4N^2}-\frac{1}{2}V(\phi_1)+P=0, \label{ekg1}\\
\rm\frac{\ddot{\phi_1}\dot{\phi_1}}{N^2}+\frac{3\dot{\Omega}\dot{\phi_1}^2}{N^2}-\frac{\dot{N}\dot{\phi_1}^2}{N^3}+\dot V(\phi_1) = 0, \label{ekg2}\\
\rm\frac{\ddot{\phi_2}\dot{\phi_2}}{N^2}+\frac{3\dot{\Omega}\dot{\phi_2}^2}{N^2}-\frac{\dot{N}\dot{\phi_2}^2}{N^3}=0,\label{ekg3}\\
\rm \rho=\rho_\gamma e^{-3(\gamma+1)\Omega} \label{e-density},
\end{align}
where $\rho_\gamma$ is an integration constant that will depend on
the epoch of the Universe being analysed.
  From the last
equation we can obtain the solution for the scalar field $\phi_2$
(in quadrature form), giving
\begin{equation}
\rm \phi_2=\phi_{2_0}+k_1 \int e^{-3\Omega} Ndt, \label{sol-phi2}
\end{equation}
where $\rm \phi_{2_0}$ and $\rm k_1$ are integration constants.

 Taking the metric
(\ref{frw1}), the Ricci scalar takes the form $\rm R=-6\frac{\ddot
\Omega}{N^2}-12 \frac{\dot \Omega^2}{N^2}+\frac{\dot \Omega \dot
N}{N^3}$ and after pluging it into (\ref{first-lagra}), where we
take the particular scalar potential $\rm V(\phi_1)=V_1\,
e^{-\lambda_1 \phi_1}$, the Lagrangian density becomes (we droped a
total time derivative, where a second time derivative of scale
factor $\Omega$ was included)

\begin{equation}\label{lagrafrw-s} \rm {\cal{L}}= \rm
e^{3\Omega}\left(\frac{6\dot{\Omega}^2}{N}-\frac{\dot{\phi_1}^2}{2N}-\frac{\dot{\phi_2}^2}{2N}
+ N V_1 e^{-\lambda_1 \phi_1} +2N\rho  \right)\,,
\end{equation}
where a $``\cdot"$ represents a time derivative.

 The associated conjugate momenta can be calculated in ordinary fashion,
that is $\partial\mathcal L/\partial\dot q_i$, which gives
\begin{equation}\label{momenta-s}
\begin{split}
\rm \Pi_\Omega &= \rm 12 \frac{e^{3\Omega}}{N}\dot \Omega, \\
\rm \Pi_{\phi_1}&=\rm  -\frac{e^{3\Omega}}{N}\dot\phi_1,\\
\rm \Pi_{\phi_2}&= \rm -e^{3\Omega} \dot \phi_2, \\
\end{split}
\qquad
\begin{split}
\rm \dot\Omega&=\rm \frac{N e^{-3\Omega}}{12} \Pi_\Omega, \\
\rm \dot \phi_1&=\rm -N e^{-3\Omega} \Pi_{\phi_1}, \\
\rm \dot \phi_2&=\rm -N e^{-3\Omega} \Pi_{\phi_2}.
\end{split}
\end{equation}
Writing (\ref{lagrafrw-s}) in a canonical form, {\it i.e.} $\rm
{\cal L}_{can}=\Pi_q\dot q-N\mathcal H$, we can perform the
variation of this canonical Lagrangian with respect to the lapse
function $N$, $\delta\mathcal L_{can}/\delta N=0$, resulting in the
constraint $\mathcal H=0$, then, the lapse function acts as a
Lagrange multiplier, hence the Hamiltonian density is
\begin{equation}
\rm {\cal H}= \frac{e^{-3\Omega}}{24} \left[ \Pi_\Omega^2-12
 \Pi_{\phi_1}^2 - 12
\Pi_{\phi_2}^2 -U(\phi_1,\Omega)-48\rho_\gamma
e^{-3(\gamma-1)\Omega} \right], \label{first-hamifrw}
\end{equation}
where $\rm U(\phi_1,\Omega)=24V_1 e^{-\lambda_1\phi_1+6\Omega}$.
The
corresponding Hamilton equations are
\begin{equation}\label{HE-all}
\begin{split}
\rm \dot{\Omega} &= \rm\frac{e^{-3\Omega}}{24}2\Pi_{\Omega},\\
\rm \dot{\phi_1} &= -\rm\frac{e^{-3\Omega}}{24}24\Pi_{\phi_1},\\
\rm \dot{\phi_2} &=-\rm\frac{e^{-3\Omega}}{24}24\Pi_{\phi_2},\\
\end{split}
\qquad
\begin{split}
\rm\dot{\Pi}_{\Omega}&=\rm\frac{e^{-3\Omega}}{24}\left[6U(\phi_1,\Omega)-144\rho_\gamma (\gamma-1)e^{-3(\gamma-1)\Omega}\right] ,\\
\rm\dot{\Pi}_{\phi_1}&=-\rm\frac{e^{-3\Omega}}{24}\lambda_1 \,U(\phi_1,\Omega),\\
\rm\dot{\Pi}_{\phi_2}&=0.
\end{split}
\end{equation}
This last set of equations cannot be decoupled due to the presence of the factor $\rm
e^{-3\Omega}$. Taking the advantage that the gauge of $\rm N$ can be fixed, dictated by the form of the Hamiltonian density
Eq.(\ref{first-hamifrw}), we can set $\rm N=24 e^{3\Omega}$, enabling us to find solutions to the problem at hand.
Now the metric (\ref{frw1}) takes the form
\begin{equation}
\rm ds^2=e^{2\Omega}\left[ -576e^{4\Omega}dt^2 +dr^2+r^2\left(d\theta^2+sin^2\theta d\phi^2\right)\right].\label{16}
\end{equation}
It is worth mentioning that applying the transformation $\rm
d\tau=24 e^{2\Omega}dt$ in the FRW metric (in Minkowskian
coordinates) one obtains a metric in conformal coordinates given by
(\ref{16}).

Working with the Hamilton's equations of motion we have that the
canonical velocities and momenta are
\begin{equation}\label{HE}
\begin{split}
\rm \dot{\Omega} &=\rm 2\Pi_{\Omega},\\
\rm \dot{\phi_1} &= \rm-24\Pi_{\phi_1},\\
\rm \dot{\phi_2} &=\rm-24\Pi_{\phi_2},\\
\end{split}
\qquad
\begin{split}
\dot{\Pi}_{\Omega}&=\rm6U(\phi_1,\Omega)-144\rho_\gamma (\gamma-1)e^{-3(\gamma-1)\Omega} ,\\
\dot{\Pi}_{\phi_1}&=\rm-\lambda_1 \,U(\phi_1,\Omega),\\
\dot{\Pi}_{\phi_2}&=\rm0.
\end{split}
\end{equation}
In the following sections we will solve this set of equations for particular values of $\gamma$.

\subsection{Master equation for $\rho_\gamma=0$ or $\rho_\gamma\not=0$ ($\gamma=1$)}
In this section we construct a master equation from Eqs.(\ref{HE})
that will allow us to obtain the different solutions for the model
under study. We start by realizing that from the last equation of
(\ref{HE}) it follows that $\rm \Pi_{\phi_2}=p_{\phi_2}=constant$.
Also, a relation between $\rm \Pi_\Omega$ and $\Pi_{\phi_1}$ can be
achieved noticing that
\begin{equation}
\rm
\frac{\dot{\Pi}_{\Omega}}{\dot{\Pi}_{\phi_1}}=-\frac{6}{\lambda_1}
\,,
\end{equation}
yielding
\begin{equation}\label{Piphi-PiOmega}
\rm \Pi_{\phi_1}=-\frac{\lambda_1}{6}\Pi_{\Omega}+p_{\phi_1} \,,
\end{equation}
where $\rm p_{\phi_1}$ is an integration constant and remains a free
parameter of the model to be adjusted with the data collected from
the cosmological observations. Also from Eqs.(\ref{HE}) we find the
following general relation between the coordinates fields $\rm
(\Omega,\phi_1)$ as follow: substituting equation
(\ref{Piphi-PiOmega}) into $\dot \phi_1$  we have $\rm \dot
\phi_1=4\lambda_1 \Pi_\Omega -24 p_{\phi_1}$, reinserting the
equation for $\dot \Omega$ and integrating, we obtain
\begin{equation}
 \rm \Delta
\phi_1= 2\lambda_1 \Delta \Omega - 24 p_{\phi_1} \Delta t.
\label{generalrelation}
\end{equation}
 On the other hand, for the case considered here, we can rewrite
the Hamiltonian (\ref{first-hamifrw}) in the following form
\begin{equation}
\rm \Pi_\Omega^2-12
 \Pi_{\phi_1}^2 - 12
\Pi_{\phi_2}^2 -48\rho_1=U(\phi_1,\Omega),
\end{equation}
and after replacing Eq.(\ref{Piphi-PiOmega}), the expression for the momenta $\rm \Pi_\Omega$ and  $\rm \Pi_{\phi_2}$, we have
\begin{equation}
\rm 2\left(3-\lambda_1^2\right)\Pi_\Omega^2 + 24\lambda_1 p_{\phi_1}
\Pi_\Omega - 72\left[p_{\phi_1}^2+ p_{\phi_2}^2+4\rho_1 \right]=
\dot \Pi_\Omega,
\end{equation}
enabling us to obtain a temporal dependence for $\rm \Pi_\Omega(t)$
which allows us to construct a master equation:
\begin{equation}
\rm \frac{d \Pi_\Omega}{a_1 \Pi_\Omega^2 + a_2 \Pi_\Omega - a_3}=dt
\,, \label{master-equation}
\end{equation}
where the parameters $\rm a_i\,,\,i=1,2,3$, are
\begin{equation}\label{parameter}
\rm a_1=2\left(3-\lambda_1^2\right) \,,\quad a_2=24\lambda_1
p_{\phi_1} \,, \quad a_3=72\left[p_{\phi_1}^2+
p_{\phi_2}^2+4\rho_1\right] \,.
\end{equation}
Subsequently by analyzing the parameter $\rm\lambda_1^2$ we will
obtain three different solutions. The first of them will be considering $\lambda_1^2<3$,
 where in \cite{OSR} according to Planck data \cite{Planck}, it has been pointed out that
 this value gives an inflationary period for a single scalar field cosmology. Secondly,
 we are going to consider the value $\lambda_1^2>3$, although the references mention
 this choice does not exhibit an inflationary epoch for a single scalar field,
 but things might change under the considered model, in light that two scalar
 fields are taken into account. Lastly, the value for $\lambda_1=\sqrt{3}$ will lead us to the third solution.
 The particular case that $\rm p_{\phi_2}=0$ was presented in \cite{OSR}, for $\lambda_1<\sqrt{3}$,
 where the model was tested with the Planck data and the analysis for the case of $\lambda_1=\sqrt{3}$
 was addressed in \cite{soj}. Also, in \cite{soco2}, using a different class of scalar potential (a product of exponential functions)
 the authors obtain exact solutions for a FRW multi-field cosmological model.

\subsubsection{Solution for $\lambda_1^2<3$.}\label{sol_lambda-3}
For this particular case, we have the following solution
\begin{equation}\label{general-solution-of-Pi-Omega}
\rm \frac{1}{24 \omega}\, Ln\left[\frac{\eta \Pi_{\Omega}+6\lambda_1
p_{\phi_1}-6 \omega} {\eta \Pi_{\Omega} +6\lambda_1
p_{\phi_1}+6\omega} \right]=t-t_0\,,
\end{equation}
where $\eta=3-\lambda_1^2 >0$, $\omega=\sqrt{3p_{\phi_1}^2+ \eta
(p_{\phi_2}^2+4\rho_1)}$ and $\rm t_0$ is a time-like integration
constant. With Eq.(\ref{general-solution-of-Pi-Omega}), the
canonical momentum $\Pi_{\phi_1}$ and the rest of variables can be
solved, hence the solutions are
\begin{eqnarray}
 \rm \Omega(t) &=& \rm \Omega_{0}-\frac{12\lambda_1
p_{\phi_1}}{\eta}(t-t_0)-\frac{1}{\eta}\,
Ln\left[Sinh\left[12\omega(t-t_0)\right]\right] \,,\label{Omegat}\\
\rm \phi_1(t) &=& \rm \phi_{1_{0}}-\frac{72
p_{\phi_1}(t-t_0)}{\eta}-\frac{2\lambda_1}{\eta}\,
Ln\left[Sinh\left[12\omega(t-t_0)\right]\right] \label{phit}\,,\\
\rm \phi_2(t) &=& \rm \phi_{2_{0}}-24 p_{\phi_2}(t-t_0), \label{phit2}\\
\rm \Pi_{\Omega}(t) &=& \rm -\frac{6\lambda_1 p_{\phi_1}}{\eta}-\frac{6\omega }{\eta}\,Coth\left[12\omega (t-t_0)\right] \,,\\
\rm \Pi_{\phi}(t) &=& \rm \frac{3p_{\phi_1}}{\eta}+\frac{\lambda_1 \omega}{\eta}\,Coth\left[12\omega(t-t_0)\right] \,,
\end{eqnarray}
where  ($\rm\Omega_0,\phi_{1_{0}},\phi_{2_{0}}$) are integration
constants and the constraint $2\eta
V_0=\omega^2\,e^{\lambda_1\phi_{1_0}-6\Omega_0}$, which is obtained
when we introduce all solutions in the set of EKG equations
(\ref{ekg0})-(\ref{ekg3}). This set of solutions is a complete and
exact classical representation of a canonical scalar field with
exponential potential in a flat FRW metric, then the scale factor
becomes
\begin{equation}\label{scale-factor-general}
\rm A(t)=A_{0}\exp\left[-\frac{12\lambda_1
p_{\phi_1}}{\eta}(t-t_0)\right]\left(Csch\left[12\omega(t-t_0)\right]\right)^{\frac{1}{\eta}},
\end{equation}
being $\rm A_{0}=e^{\Omega_{0}}$. From (\ref{scale-factor-general})
it is evident that the scale factor has a decreasing behavior,
therefore, in order to have a growing volume function in the
inflationary epoch we must impose that $\rm p_{\phi_1}<0$
(which must be taken into account in all the equations in this section). As we will see below, the new sign will be reflected
in the deceleration parameter. One may question if there is a
procedure to rewrite Eq.~(\ref{scale-factor-general}) in terms of
$\rm t_{phys}$, where $\rm t_{phys}=\int dt N(t)$, however, a
forthright relation between $\rm t_{phys}$ and t is far from being
determined, since one must first compute such integral, if possible,
and then obtain $\rm t=t(t_{phys})$ which can be a nontrivial
endeavor. However, all observable parameters must be evaluated at
$\rm t_{phys}$, or in terms of an equivalent evolution variable, yet
ascertaining an appropriate manipulation of the gauge.
\subsubsection{Solution for $\lambda_1^2>3.$}\label{sol_lambda+3}

 For this case $\eta <0$
and $\rm a_1=2(3-\lambda_1^2)<0$, so the master equation
(\ref{master-equation}) can be casted as
\begin{equation}\label{22}
\rm \frac{d \Pi_\Omega}{-m_1 \Pi_\Omega^2+  a_2 \Pi_\Omega -
a_3}=dt
\end{equation}
where we have included the minus sign such that the constant $\rm
m_1=2(\lambda_1^2-3)=2\beta>0$. Then, defining $\rm \omega_1^2
=a_2^2 -8\beta a_3=576 \omega_2^2$ with $\omega_2^2=3\rm
p_{\phi_1}^2-\beta( p_{\phi_2}^2 +4\rho_1)$, we can rewrite
(\ref{22}) as
\begin{equation}
\frac{8\beta \,d
\Pi_\Omega}{\omega_1^2-\left(4\beta \Pi_\Omega-24\lambda_1
\rm p_{\phi_1} \right)^2}=\rm dt,\label{master-integral-lambda-bigger3}
\end{equation}
where the constraint over the parameters $\rm
p_{\phi_1}>\sqrt{(p_{\phi_2}+4\rho_1)\left[\left(\frac{\lambda_1}{\sqrt{3}}\right)^2-1\right]}$
must be fulfilled. In order to be able to integrate
Eq.(\ref{master-integral-lambda-bigger3}), as a final step, we
resort to the change of variables $\rm z=4\beta
\Pi_\Omega-24\lambda_1 p_{\phi_1}$, thus, the solution for the
momenta $\rm \Pi_\Omega(t)$ becomes
\begin{equation}
\rm \Pi_\Omega=\frac{6\lambda_1
p_{\phi_1}}{\beta}+\frac{6\omega_2}{\beta} Tanh\left( 12 \omega_2
(t-t_0) \right) \,.
\end{equation}
Using the relations from Eq.(\ref{HE}) and after some algebra, the
solutions for the set of variables $\rm(\Omega,\phi_1, \phi_2)$ and
$\rm (\Pi_{\phi_1},\Pi_{\phi_2})$ are:
\begin{eqnarray}
&&\rm \Omega= \Omega_0 + \frac{12\lambda_1 p_{\phi_1}}{\beta}(t-t_0) + \frac{1}{\beta}\, Ln\left[Cosh\left(12\omega_2(t-t_0)\right)\right]  \,,\\
&&\rm \phi_1 =\phi_{1_{0}} + 72\frac{p_{\phi_1}}{\beta} (t-t_0) +\frac{2\lambda_1}{\beta}\,Ln\left[Cosh\left(12\omega_2(t-t_0)\right) \right] \,, \\
&&\rm \phi_2= \phi_{2_{0}} -24 p_{\phi_2}(t-t_0), \\
&&\rm \Pi_{\phi_1}=-\frac{3p_{\phi_1}}{\beta}-\frac{\lambda_1\omega_2}{\beta} Tanh\left(12\omega_2(t-t_0) \right) \,, \\
&&\rm \Pi_{\phi_2}=p_{\phi_2}\,,
\end{eqnarray}
where ($\rm\Omega_0,\phi_{1_{0}},\phi_{2_{0}}$) are all integration
constants. In order that the above solutions fulfill the EKG
Eqs.(\ref{munu}-\ref{ekg-phi2}), all constants must satisfy that
$\rm 2 \beta V_0=\omega_2^{2}e^{\lambda_1 \phi_{1_{0}}
-6\Omega_{0}}$. Finally the scale factor becomes
\begin{equation}
\rm A(t)=A_0\, Exp\left[ \frac{12\lambda_1 p_{\phi_1}}{\beta}(t-t_0)
\right] \,\,Cosh^{\frac{1}{\beta}}\left(12\omega_2(t-t_0)\right),\label{volumen2}
\end{equation}
here, as before, $\rm A_{0}=e^{\Omega_{0}}$.
\subsubsection{Solution for $\lambda_1^2=3.$}\label{sol_lambda3}

 For completeness, we include the exotic case for $\lambda_1^2=3$ (and $\eta =0$)
 which emerge as consequence of SUSY Quantum Mechanics applied to cosmological models \cite {socorro3};
 for this case the coefficient $\rm a_1=0$ and the master equation to solve is reduced to
\begin{equation}
\rm \int\frac{d \Pi_\Omega}{a_2 \Pi_\Omega - a_3} = \int dt \,,
\end{equation}
thus $\rm \Pi_\Omega(t)$ becomes
\begin{equation}\label{Pi-Omega-lambda2-equal-3}
\rm\Pi_\Omega(t)=\frac{a_3}{a_2} +p e^{a_2(t-t_0)} \,,
\end{equation}
where $\rm p$ is an integration constant. As before, we can use relations from
Eq.(\ref{HE}) and after some manipulation, the solutions for $\rm(\Omega,\phi_1, \phi_2)$ and $\rm
(\Pi_{\phi_1},\Pi_{\phi_2})$ are:
\begin{eqnarray}
&&\rm \Omega=\Omega_0 +2\sqrt{3}
\frac{(p_{\phi_1}^2+p_{\phi_2}^2+4\rho_1)}{p_{\phi_1}}(t-t_0)
+ \frac{\sqrt{3}p}{36p_{\phi_1}} e^{24\sqrt{3}p_{\phi_1}(t-t_0)} \label{32},\\
&&\rm \phi_1=\phi_{1_{0}} + 12\frac{(p_{\phi_2}^2+4\rho_1)-p_{\phi_1}^2+}{p_{\phi_1}} (t-t_0) + \frac{p}{6} e^{24\sqrt{3}p_{\phi_1}(t-t_0)} \,, \\
&&\rm \phi_2=\phi_{2_{0}}  -24 p_{\phi_2} (t-t_0) \,, \\
&&\rm \Pi_{\phi_1}=\frac{1}{2}\frac{p_{\phi_1}^2-(p_{\phi_2}^2+4\rho_1)}{p_{\phi_1}} -\frac{\sqrt{3}p}{6}  e^{24\sqrt{3}p_{\phi_1}(t-t_0)} \,, \\
&&\rm \Pi_{\phi_2}=p_{\phi_2}\label{36},
\end{eqnarray}
again ($\rm\Omega_{0},\phi_{1_{0}},\phi_{2_{0}},$) are all
integration constants. If we want the equations (\ref{32})-(\ref{36}) satisfy the EKG Eqs.(\ref{munu})-(\ref{ekg-phi2}),
all constants must fulfil that
$\rm V_0=\frac{\sqrt{3}p_{\phi_1}}{6} p\, e^{-6\Omega_0+\lambda_1
\phi_{1_{0}}}$. Finally the scale factor $\rm A(t)$ for this case is
\begin{equation}\label{huge-scale}
\rm A(t)=A_0 Exp\left[2\sqrt{3}
\frac{p_{\phi_1}^2+p_{\phi_2}^2+4\rho_1}{p_{\phi_1}}(t-t_0)\right]
\,Exp\left[ \frac{\sqrt{3}p}{36p_{\phi_1}}
e^{24\sqrt{3}p_{\phi_1}(t-t_0)}\right] \,,
\end{equation}
where $\rm A_{0}=e^{\Omega_{0}}$.
\subsubsection{Deceleration and barotropic parameters.}
\begin{figure}[ht!]
\begin{center}
\captionsetup{width=.9\textwidth}
\includegraphics[scale=0.4]{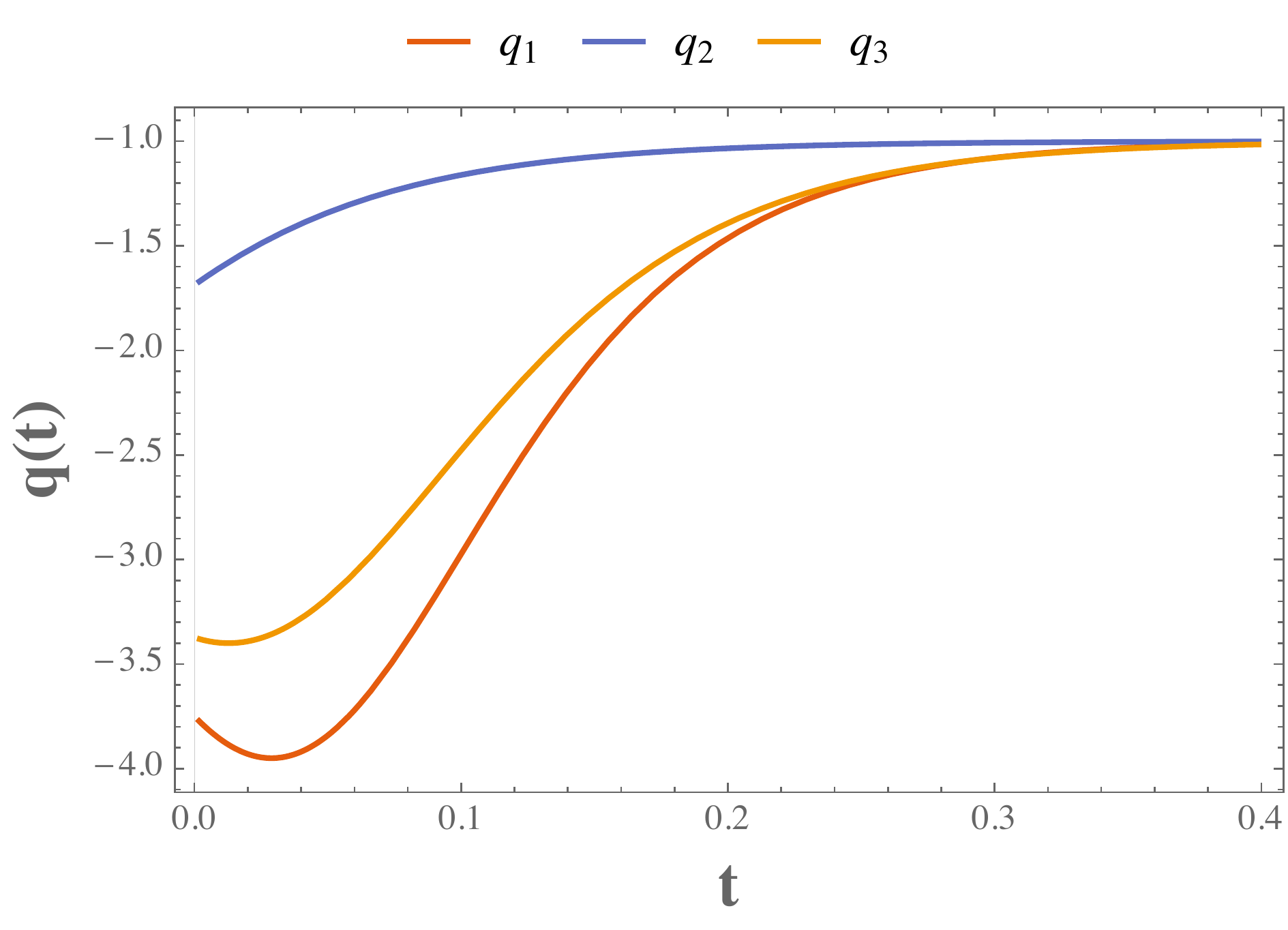}
\includegraphics[scale=0.4]{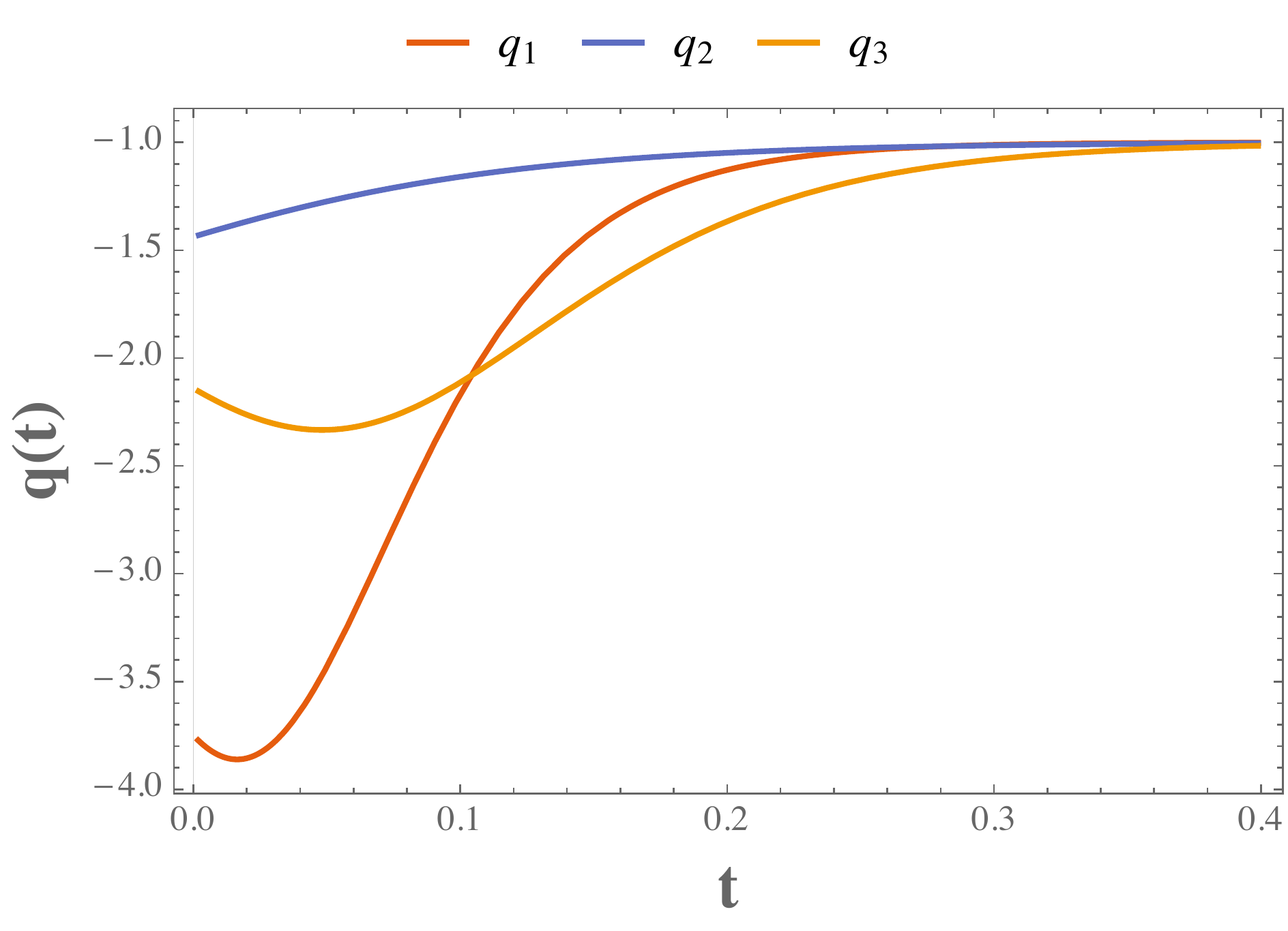}
\caption{Deceleration parameter for the three classical solutions. Here we
have taken (for both figures) $\lambda_1=0.5$, $\lambda_1^{\prime}=2$, $\rm
p_{\phi_1}=0.4$, $\rm p_{\phi_2}=0.2$ and $\rm p=0.7$. For the left figure we set $\rho_1=0$,
whereas for the right one $\rho_1=0.04$}\label{q-parameters}
\end{center}
\end{figure}

One way that we can analyze the dynamical behavior of the scale
factor for each of the three solutions of the model under
consideration, is to calculate the deceleration parameter, which is
defined as
\begin{equation}
\rm q=-\rm\frac{A\ddot A}{\dot A^2}.
\end{equation}
 The corresponding deceleration parameters are given by
\begin{align}
\rm q_1&=\rm -1-\frac{\eta\omega^2}{\left[\lambda_1p_{\phi_1}Sinh(12\omega t)-\omega Cosh(12\omega t)\right]^2},\label{q1}\\
\rm q_2&=\rm -1-\frac{\beta\omega^{\prime2}}{\left[\lambda_1^{\prime}p_{\phi_1}Cosh(12\omega^{\prime} t)+\omega^{\prime} Sinh(12\omega^{\prime} t)\right]^2},\label{q2}\\
\rm q_3&=\rm -1-\frac{12\sqrt{3}pp_{\phi_1}^3
Exp(24\sqrt{3}p_{\phi_1}t)}{[pp_{\phi_1}Exp(24\sqrt{3}p_{\phi_1}t)+\sqrt{3}(p_{\phi_1}^2+p_{\phi_2}^2+4\rho_1)]^2},\label{q3}
\end{align}
where $\rm q_1, q_2$ and $\rm q_3$ stand for the solutions for
$\lambda_1^2<3$, $\lambda_1^2>3$ and $\lambda_1^2=3$, respectively;
also
$\rm\omega^{\prime}=\sqrt{3p_{\phi_1}^2-\eta^{\prime}(p_{\phi_2}^2+4\rho_1)}$
and $\rm \eta^{\prime}=(\lambda_1^{\prime})^2-3$,
($\lambda_1^{\prime}$ being a constant). In Fig.(\ref{q-parameters})
we can see the behavior of each of the deceleration parameters when
matter is absent (left panel) and when matter is incorporated (right
panel). For the scenario $\rho_1=0$ it is observed that $q_1$
exhibits a deceleration and acceleration behavior. $q_3$ also
presents a deceleration/acceleration stage, being the former only
for a short period of time (compared to $\rm q_1$) to then
accelerate. When matter is incorporated things change for $\rm q_1$
and $\rm q_3$, from the right panel of Fig.~(\ref{q-parameters}),
 we can see that the former, again, presents a deceleration period to later have a sudden growth. In this setup, $\rm q_3$ starts
 in a slow deceleration stage to then gradually accelerate. Lastly, in contrast to $\rm q_1$ and $\rm q_3$, $\rm q_2$ has
 only an accelerated stage with the similarity that in both scenarios the growth is slow.  After a sufficient period of time the
 three solutions (in both scenarios) converge to the same value -1.
Also, we can write the barotropic parameter $\rm
\omega_{T}=\frac{(Pressure)_{total}}{(energy~density)_{total}}$ that
is related to the deceleration parameter q under the Einstein

equations (\ref{ekg0}) and (\ref{ekg1}), for the gauge $\rm
N\not=1$, as
\begin{equation}
\omega_T=\rm\frac{2}{3}q+\frac{5}{3},\label{omegat}
\end{equation}
and taking the deceleration parameters given by equations (\ref{q1})-(\ref{q3}) the barotropic parameter for each of the solutions are given by
\begin{align}\label{parametro-w}
\rm \omega_{T_1}&=\rm 1-\frac{2}{3}\frac{\eta \omega^2}{\left[\lambda_1p_{\phi_1}Sinh(12\omega t)-\omega Cosh(12\omega t)\right]^2},\\
\rm \omega_{T_2}&=\rm 1-\frac{2}{3}\frac{\beta {\omega^\prime}^2}{\left[\lambda_1^{\prime}p_{\phi_1}Cosh(12\omega^{\prime} t)
+\omega^{\prime} Sinh(12\omega^{\prime} t)\right]^2},\\
\rm \omega_{T_3}&=\rm 1-\frac{2}{3}\frac{12\sqrt{3}pp_{\phi_1}^3
Exp(24\sqrt{3}p_{\phi_1}t)}{[pp_{\phi_1}Exp(24\sqrt{3}p_{\phi_1}t)+\sqrt{3}(p_{\phi_1}^2+p_{\phi_2}^2+4\rho_1)]^2}.
\end{align}
\begin{figure}[ht!]
\begin{center}
\captionsetup{width=.8\textwidth}
\includegraphics[scale=0.6]{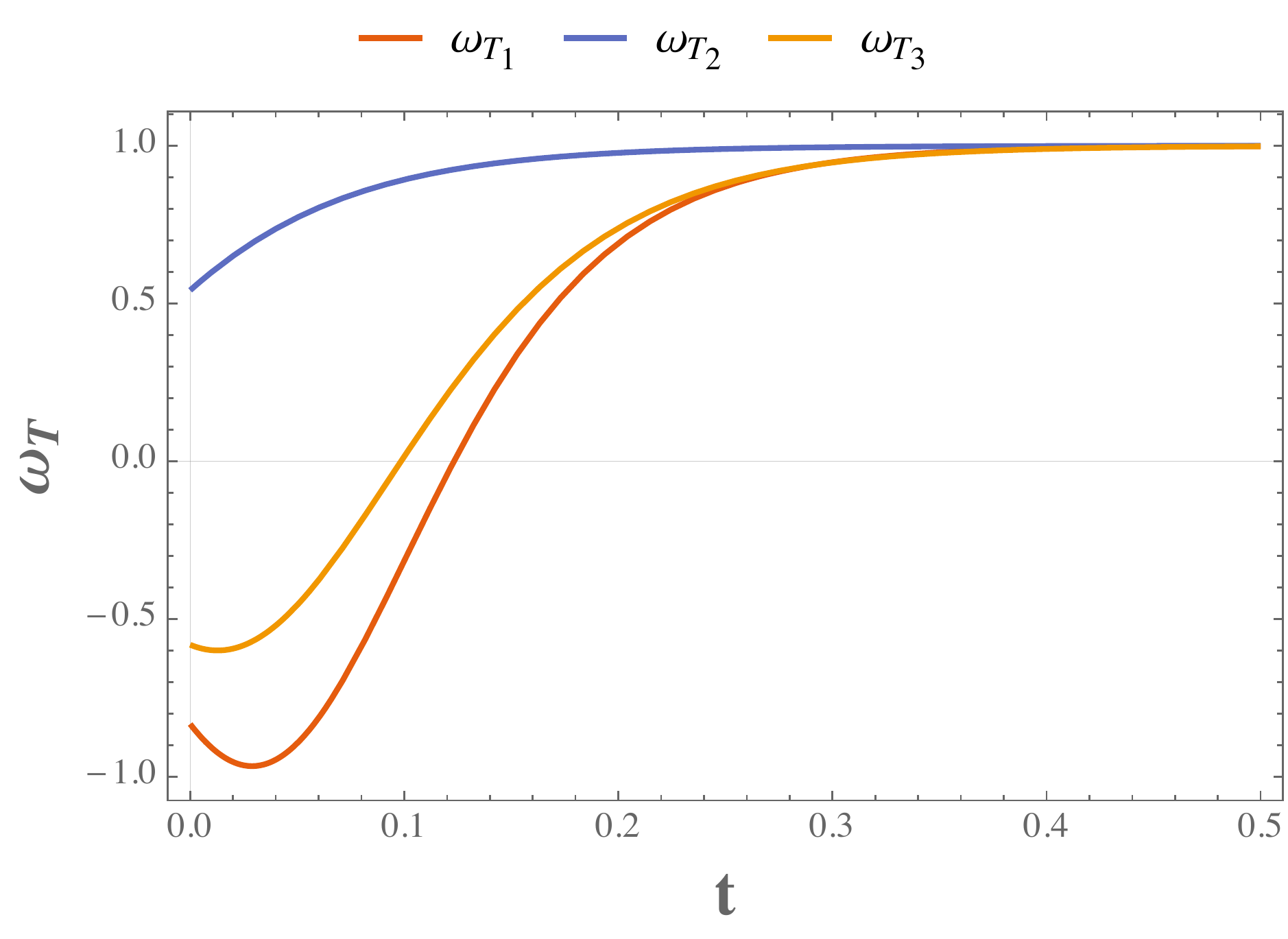}
\caption{Barotropic parameters for the three different solutions
without standard matter. For the three $\omega_T$ we have taken
$\lambda_1=0.5$, $\lambda_1^{\prime}=2$, $\rm p_{\phi_1}=0.4$, $\rm
p_{\phi_2}=0.2$ and $\rm p=0.7$.}\label{omegas-polvo}
\end{center}
\end{figure}
Analyzing the behavior $\omega_{T_i}$ from the last set of equations
it can be observed that for late times all three values tend to 1.
It is a well known fact that for a quintessence model the lower and
upper bounds are given by $\rm -1\leq\omega_T\leq 1$,
from Fig~(\ref{omegas-polvo}) we can see that our model fits within this parameters.\\
For completeness, the barotropic parameter in the gauge $\rm N=1$ becomes
\begin{equation}
\rm \omega_T=\rm \frac{2}{3}q-\frac{1}{3}.\label{norma}
\end{equation}


\subsection{Standard matter scenario ($\gamma\not=1$)}\label{2.2}
Trying to solve the set of equations (\ref{HE}) for values of
$\gamma\not=1$ under Hamilton's approach is a dead end. To sort out
this setback, we must implement another method. For this particular
case we start with arbitrary scalar potential field $\rm V(\phi_1)$
employing the gauge $\rm N=1$ and the EKG equations. As we are
considering a flat FRW Universe, the line element can be written as
\begin{equation}
\rm ds^2=- dt^2 +A^2(t) \left[dr^2 +r^2(d\theta^2+sin^2\theta
d\phi^2) \right], \label{frw11}
\end{equation}
where  $\rm A(t)$ is the scale factor of the model. The
corresponding EKG equations are
\begin{eqnarray}
\rm 3 \left(\frac{\dot A}{A}\right)^2 -\frac{1}{4} {\dot \phi_1}^2
-\frac{1}{4}{\dot \phi_2}^2 -8\pi G_N \rho -\frac{1}{2}V(\phi_1)=0,\label{freeman}\\
\rm 2\frac{\ddot A}{A}+\left(\frac{\dot A}{A}\right)^2 +\frac{1}{4}
{\dot \phi_1}^2 +\frac{1}{4}{\dot \phi_2}^2 +8\pi G_N \gamma \rho-\frac{1}{2}V(\phi_1)=&0,\label{cons}
\end{eqnarray}
\begin{align}
\rm -3\frac{\dot A}{A}{\dot \phi_1}^2 -\ddot \phi_1 \dot\phi_1 -\frac{\partial V(\phi_1)}{\partial t}&=\rm 0, \quad \to \quad \frac{d}{dt}Ln\left(A^3 {\dot \phi_1}\right)=-\frac{\dot V}{{\dot \phi_1}^2}, \label{phi11}\\
\rm 3\frac{\dot A}{A}\dot \phi_2 + \ddot \phi_2=\rm & \rm 0, \quad \to\qquad \Delta \phi_2 =c_2 \int \frac{dt}{A^3}. \label{phi22}
\end{align}
Considering a barotropic equation of state for the scalar field of
the form $\rm P_\phi=\omega_\phi \rho_\phi$ we can obtain the
kinetic energy $\rm K_\phi= \frac{1+\omega_\phi}{1-\omega_\phi}
V(\phi)$, therefore, equation (\ref{phi11}) can be written in
general way as
\begin{equation}
\rm \frac{d}{dt} \left[ Ln\left(A^6 V^{\frac{2}{1+\omega_\phi}}\right) \right]=0  \quad\rightarrow \quad V=c_\omega A^{-3(1+\omega_\phi)}, \label{potential}
\end{equation}
hence, solutions to the scalar field can be expressed in quadrature form as
\begin{equation}
\rm \Delta \phi_1 = \alpha_\omega \int
\frac{dt}{A^{\frac{3(1+\omega_\phi)}{2}}}. \label{scalar-field}
\end{equation}
For the attractor case $\omega_\phi=\gamma\not=1$, Eq.(\ref{freeman}) can be casted as
\begin{equation}
\rm  3 \left(\frac{\dot A}{A}\right)^2 = \frac{c_1^2+8\pi G\rho_\gamma}{A^{3(1+\gamma)}} + \frac{c_2^2}{A^6}, \quad \to \quad
\frac{A^2\, dA}{\sqrt{b_\gamma A^{3(1-\gamma)} + b_2}}=dt,
\label{mast}
\end{equation}
where we have identified the scalar fields contributions as
$\rm\frac{1}{4}{\dot \phi_2}^2=\frac{c_2^2}{A^6}$ and
$\rm\frac{1}{4}{\dot \phi_1}^2=\frac{c_1^2}{A^{3(1+\omega_\phi)}}$
with $\rm c_1$ and $\rm c_2$ integration constants (see equations
(\ref{phi22}) and, (\ref{potential})); and $\rm
b_\gamma=\frac{c_1^2}{3}+\frac{8}{3}\pi G \rho_\gamma$ and $\rm
b_2=\frac{c_2^2}{3}$.
The general solution of (\ref{mast}) is given in terms of a
Hypergeometric function,
\begin{equation}\label{A_hiper}
\rm  \Delta t =\frac{A^3}{\sqrt{3}c_2} \, _2F_1\left[\frac{1}{2},
\frac{1}{1-\gamma},\frac{2-\gamma}{1-\gamma}, -\frac{8\pi G
\rho_\gamma+c_1^2}{c^2}A^{3(1-\gamma)} \right],\quad(\gamma\neq1).
\end{equation}
\subsubsection{Dust Epoch.}
\begin{figure}[ht!]
\begin{center}
\includegraphics[scale=0.7]{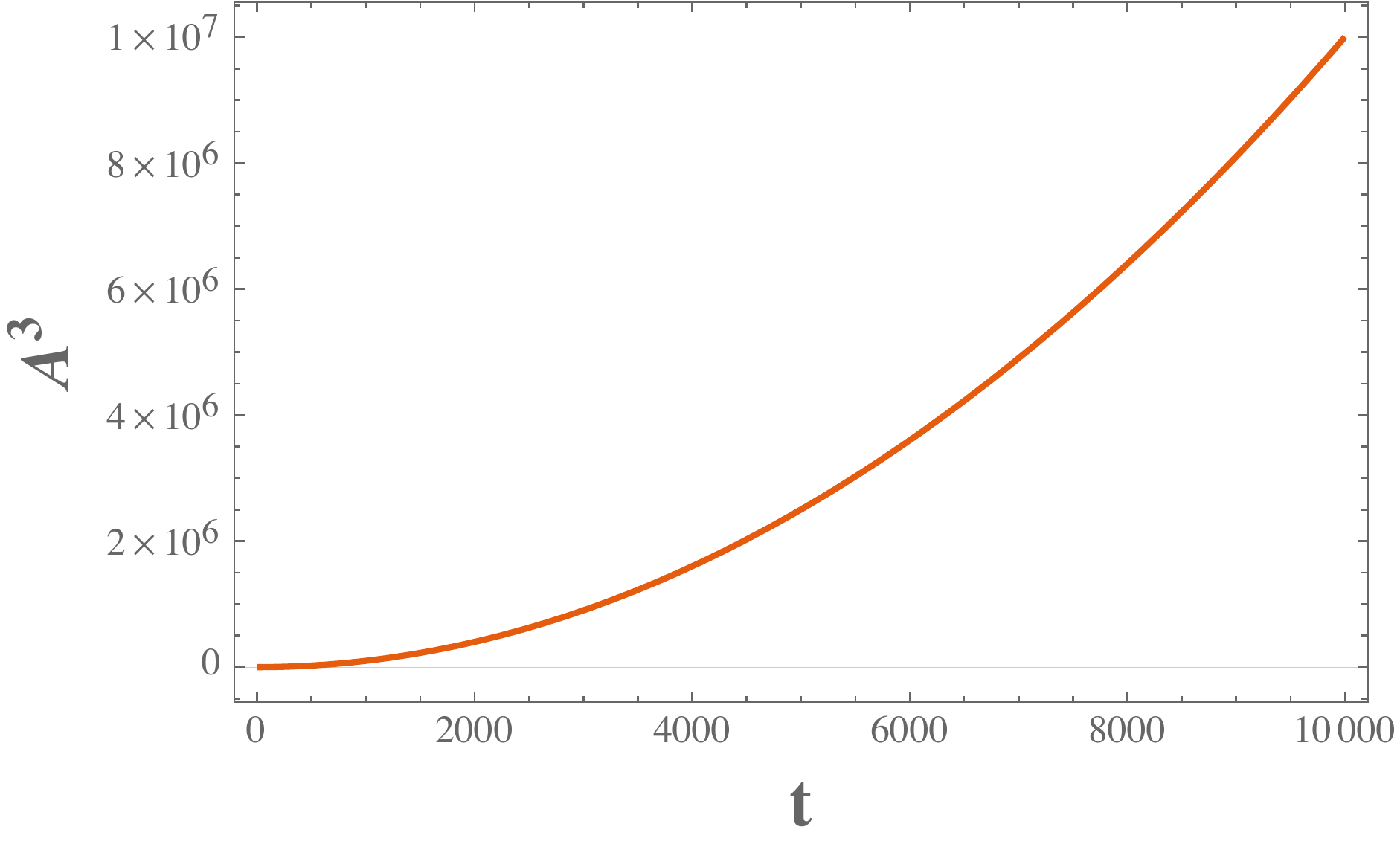}
\end{center}
\caption{Volume function for the dust epoch in the standard matter,
for $\rm a_0=0.001$, $\rm a1=a_2=10$.}\label{volumen_polvo}
\end{figure}

For the dust scenario in the standard matter we choose $\gamma=0$,
therefore equation (\ref{mast}) reduces to
\begin{equation}
\rm  \frac{A^2\, dA}{\sqrt{b_0 A^3 + b_2}}=dt,
\end{equation}
being $\rm b_0=\frac{c_1^2}{3}+\frac{8}{3}\pi G \rho_0$ and $\rm
b_2=\frac{c_2^2}{3}$. Considering the change of variables $\rm z=b_0
A^3 + b_2$ we arrived to the solution for the scale factor, which is
\begin{equation}
\rm A^3(t)=a_0\left[(a_1 t +a_2)^2-1\right],
\label{solution-a}
\end{equation}
here $\rm a_0=\frac{b_2}{b_0}=\frac{c_2^2}{c_1^2+8\pi G \rho_0}$, and $\rm a_1=\frac{3b_0}{2\sqrt{b_2}}=
\frac{\sqrt{3}}{2}\frac{c_1^2+8\pi G\rho_0}{c_2}$, and its dynamical behavior can be observed in Fig.(\ref{volumen_polvo}).
Now with this solution at hand, we can write the scalar fields as
\begin{align}\label{sol-phi11}
\rm \phi_1(t)&=\rm \phi_{1_0}+\frac{c_1}{a_1}\sqrt{\frac{2}{a_0}}Ln\left[u+\sqrt{u^2-1}\right]\nonumber\\
&=\rm \phi_{1_0}+\frac{4}{\sqrt{6(c_1^2+8\pi G \rho_0)}}\,Ln\left[\rm u+\sqrt{\rm u^2-1}\right],\\
\rm \phi_2(t)&= \rm\phi_{2_0}+\frac{2}{\sqrt{3}}Ln\left(1-\frac{2}{u+1}\right), \label{sol-phi22}
\end{align}
where $\rm u(t)=a_1t+a_2$. From (\ref{sol-phi11}) is possible to rewrite the variable $\rm u$ in terms of the $\phi_1$ as $\rm
u=Cosh\left(\frac{\sqrt{6}}{4}\sqrt{c_1^2+8\pi G \rho_0} \Delta\phi_1 \right)$, so the corresponding scalar potential becomes
\begin{equation}
\rm V(\phi_1)=\frac{c_1^2(c_1^2+8\pi G \rho_0)}{c_2^2} Csch^2\left(\frac{\sqrt{6}}{4}\sqrt{c_1^2+8\pi G \rho_0}
\Delta \phi_1\right).\label{potential_phi}
\end{equation}
At this point it is worth mentioning that this kind of potential is
consistent with a (volume) accelerated expansion in the dust
scenario. Also, from (\ref{sol-phi22}) we can see that the scalar
field $\phi_2$ acquires a constant value of  $\phi_{2_0}$ for late
times. In a previous work \cite{los4}, this type of scalar potential
was found for an anisotropic cosmological model were the anisotropic
functions vanish for late times.

With this results we can analyze the dynamical behavior of the
volume $\rm v=A^3$ calculating the deceleration parameter, which for
the present setup takes the following form
\begin{equation}
\rm q= -\frac{v\ddot v}{{\dot v}^2},
\end{equation}
resulting in
\begin{equation}\label{q-gama-difcero}
\rm q(t)=\rm -\frac{1}{2} +\frac{1}{2u^2}.
\end{equation}
We can easily check that taken the limit $\rm t\to\infty$ in
(\ref{q-gama-difcero}) the deceleration parameter is $\rm
q(t)=-\frac{1}{2}$ (remember that $\rm u$ is a function that depends
linearly with respect to $\rm t$), indicating us that the Universe
undergoes a volume accelerated expansion, supporting the above
results. In \cite{los4} we found similar behavior for the volume
function but for an anisotropic cosmology. Also, for this case the
barotropic parameter $\omega_T$, is
\begin{equation}\label{omega-difcero}
\rm\omega_T=\frac{2}{3}q-\frac{1}{3}=\frac{1}{3u^2}-\frac{2}{3},
\end{equation}
we can observe that the asymptotic behavior of $\omega_T$ (with
respect to time) tends to $-\frac{2}{3}$ which is a signal that the
volume function has a big expansion.
\subsection{Radiation epoch ($\gamma=\frac{1}{3}$)}
Finally, we address the radiation epoch where $\gamma=\frac{1}{3}$.
For this scenario the Friedmann equation (\ref{mast}) displays the
following form
\begin{equation}
\rm \frac{A^2\, dA}{\sqrt{b_0 A^2 + b_2}}=dt,
\end{equation}
and whose solution is given by
\begin{equation}\label{solucion_radiacion}
\rm 2b_0\,\Delta t=A\sqrt{b_0\,A^2
+b_2}-\frac{b_2}{\sqrt{b_0}}\, Ln\left[A+\sqrt{A^2+\frac{b_2}{b_0}}\right].
\end{equation}
Unfortunately we don't have enough information to be able to write
the scale factor as a function of time, however, one would expect
that by inverting equation (\ref{solucion_radiacion}) the functional
form of $\rm A^3(t)$ would be an increasing function.
\section{Second Model}\label{model-two}
The next cosmological model to consider is one where in addition to
considering the two scalar fields, two potential terms also come
into play. The action for such a Universe is
\cite{chervon1995,Chervon2013,Chervon2015,Fomin2017,Paliathanasis2019}
\begin{equation}
\rm {\cal L}=\sqrt{-g} \left( R-\frac{1}{2}g^{\mu\nu}
m^{ab}\nabla_\mu \phi_a \nabla_\nu \phi_b  + V(\phi_1,\phi_2)\right)
\,, \label{lagra}
\end{equation}
where $\rm R$ is the Ricci scalar, $\rm V(\phi_1,\phi_2)$ is the
corresponding scalar field potential, and $\rm m^{ab}$ is a
 $2 \times 2$ constant matrix and $\rm m^{12}=m^{21}$. 
The corresponding variations of Eq.(\ref{lagra}), with respect
to the metric and the scalar fields gives the EKG field equations
\begin{equation}
\rm G_{\alpha \beta}=\rm -\frac{1}{2}m^{ab} \left(\nabla_\alpha
\phi_a \nabla_\beta \phi_b -\frac{1}{2}g_{\alpha \beta} g^{\mu \nu}
\nabla_\mu \phi_a \nabla_\nu \phi_b \right) +\frac{1}{2}g_{\alpha \beta} \, V(\phi_1,\phi_2), \label{mono}
\end{equation}
\begin{equation}
\rm m^{ab}\Box \phi_b -\frac{\partial V}{\partial \phi_a} =\rm
m^{ab}g^{\mu\nu} {\phi_b}_{,\mu\nu} - m^{ab}g^{\alpha \beta}
\Gamma^\nu_{\alpha \beta} \nabla_\nu \phi_b - \frac{\partial
V}{\partial \phi_a}=\rm 0, \,\label{ekg-phi}
\end{equation}
where $a,b=1,2$. The line element to be considered for this two-field cosmological model is the flat FRW
\begin{equation}
\rm ds^2=-N(t)^2 dt^2 +e^{2\Omega(t)} \left[dr^2
+r^2(d\theta^2+sin^2\theta d\phi^2) \right], \label{frw}
\end{equation}
here, as in the previous case, $\rm N$ represents the lapse
function, $\rm A(t)=e^{\Omega(t)}$ the scale factor in the Misner
parametrization and  $\rm \Omega$ a scalar function whose interval
is $ (-\infty,\infty)$. Consequently the Klein-Gordon equations are
(here the $\prime =\frac{d}{d\tau}, d\tau=Ndt$)
\begin{equation}
\rm m^{11}{\phi_1^{\prime \prime}}{\phi_1^\prime}+
m^{12}{\phi_2^{\prime \prime}}{\phi_1^\prime}  +3
\Omega^\prime\left(m^{11}{\phi_1^\prime}^2 + m^{12}  \phi_1^\prime
\phi_2^\prime \right) +\left(\dot V\right)_{\phi_2} = 0 \,,
\label{ein2}
\end{equation}
\begin{equation}
\rm m^{22}{\phi_2^{\prime \prime}}{\phi_2^\prime}+
m^{12}{\phi_1^{\prime \prime}}{\phi_2^\prime} +3
\Omega^\prime\left(m^{22}{\phi_2^\prime}^2 + m^{12}  \phi_1^\prime
\phi_2^\prime \right)+\left(\dot V\right)_{\phi_1} = 0. \,
\label{ein3}
\end{equation}
where $\left(\dot V\right)_{\phi_i}$ means that the time derivative
is calculated maintaining $\phi_i$ constant. Building the
corresponding Lagrangian and Hamiltonian densities for this
cosmological model, classical solutions to EKG
Eqs.(\ref{mono}-\ref{ekg-phi}) can be found using the Hamilton's
approach; and also the quantum formalism can be determined and
solved, as we will show below. In this line of thought, plugging the
metric Eq.(\ref{frw}) into Eq.(\ref{lagra}) and where we have taken
the particular scalar potential $\rm V(\phi_1,\phi_2)=V_1\,
e^{-\lambda_1 \phi_1}+V_2\, e^{-\lambda_2 \phi_2}$, which is
appropriate for this model, now the Lagrangian density reads
\begin{equation}\label{lagrafrw}
\rm {\cal{L}}= \rm
e^{3\Omega}\left(\frac{6\dot{\Omega}^2}{N}-\frac{m^{11}\dot{\phi_1}^2}{2N}-\frac{m^{22}\dot{\phi_2}^2}{2N}
-\frac{m^{12}\dot{\phi_1} \dot \phi_2}{N}+ N V_1 e^{-\lambda_1
\phi_1} +N V_2 e^{-\lambda_2 \phi_2} \right)\,,
\end{equation}
here, as before, an upper $``\cdot"$ represents a time derivative.
In (\ref{lagrafrw}) we can also include a contribution term $\rm
2N\rho$, regarding the standard matter content, as it was done in
the Lagrangian density (\ref{lagrafrw-s}); for the case of stiff
matter ($\gamma=1$), this contribution in the Hamiltonian density
would be reflected with the addition of a constant term and the
treatment could be carried out in the same manner as in
section~\ref{model}, however, in the following we will perform an
analysis without standard matter. From (\ref{lagrafrw}) the
resulting momenta are given by
\begin{equation}\label{mom_conjugados_26}
\begin{split}
\rm \Pi_\Omega &= \rm 12 \frac{e^{3\Omega}}{N}\dot \Omega,\\
\rm \Pi_{\phi_1}&=\rm  -\frac{e^{3\Omega}}{N}\left( m^{11}\dot\phi_1 +m^{12} \dot \phi_2\right),\\
\rm \Pi_{\phi_2}&=\rm -\frac{e^{3\Omega}}{N}\left( m^{12}\dot\phi_1 +m^{22} \dot \phi_2\right),
\end{split}
\qquad
\begin{split}
\rm \dot \Omega&=\rm \frac{N e^{-3\Omega}}{12} \Pi_\Omega,\\
\rm \dot \phi_1&=\rm N \frac{e^{-3\Omega}}{\triangle}\left(-m^{22} \Pi_{\phi_1}+m^{12} \Pi_{\phi_2} \right),\\
\rm \phi_2&=\rm N\frac{e^{-3\Omega}}{\triangle}\left(m^{12} \Pi_{\phi_1}- m^{11}\Pi_{\phi_2} \right),
\end{split}
\end{equation}
where $\triangle =m^{11}m^{22}-\left(m^{12}\right)^2$.
 Writing (\ref{lagrafrw}) in a canonical form, {\it i.e.} $\partial\mathcal L_{can}=\Pi_q\dot q-N\mathcal H$,
 we can perform the variation of this canonical Lagrangian with respect to the lapse function
 $N$, $\delta\mathcal L_{can}/\delta N=0$, resulting in the constraint $\mathcal H=0$, hence the Hamiltonian density is
\begin{eqnarray} \rm {\cal H}&=&\rm  \frac{e^{-3\Omega}}{24} \left[
\Pi_\Omega^2-\frac{12 m^{22}}{\triangle} \Pi_{\phi_1}^2-\frac{12
m^{11}}{\triangle}\Pi_{\phi_2}^2+\frac{24 m^{12}}{\triangle}
\Pi_{\phi_1}\Pi_{\phi_2}\right.\nonumber\\
&&\rm \left. -24V_1 e^{-\lambda_1\phi_1+6\Omega} -24V_2
e^{-\lambda_2\phi_2+6\Omega}\right] \,. \label{hamifrw}
\end{eqnarray}
Proposing the following canonical transformation on the variables
$\rm(\Omega,\phi_1,\phi_2)\leftrightarrow (\xi_1,\xi_2,\xi_3)$
\begin{equation}\label{trans_2}
\begin{split}
\rm \xi_1&=-6\Omega+\lambda_1 \phi_1,\\
\xi_2&= -6 \Omega+\lambda_2 \phi_2,\\
 \xi_3&=-4 \Omega + \frac{\lambda_1}{6}
\phi_1 + \frac{\lambda_2}{6} \phi_2,
\end{split}
\quad\longleftrightarrow\quad
\begin{split}
\Omega&=\rm\frac{\xi_1 + \xi_2- 6 \xi_3}{12},\\
\rm \phi_1&= \rm \frac{3\xi_1 + \xi_2-6\xi_3}{2\lambda_1},\\
\rm \phi_2 &= \rm \frac{\xi_1+3\xi_2-6\xi_3}{2\lambda_2},
\end{split}
\end{equation}
and setting the gauge $\rm N=24e^{3\Omega}$, allows us to find a new set of conjugate momenta $\rm (P_1,P_2,P_3)$
\begin{align}
\rm \Pi_\Omega &= \rm -6 P_1 -6 P_2-4 P_3, \nonumber\\
\rm \Pi_{\phi_1} &= \rm \lambda_1 P_1 +  \frac{\lambda_1}{6} P_3,
\nonumber\\
\rm \Pi_{\phi_2} &= \rm \lambda_2 P_2 +   \frac{\lambda_2}{6} P_3,
\label{new-moment}
\end{align}
which finally leads us to the Hamiltonian density
\begin{align}
\rm {\cal H} &= \rm  12 \left(3-\frac{\lambda_1^2
m^{22}}{\triangle}\right)P_1^2+12\left(3-\frac{\lambda_{2}^2\*m^{11}}{\triangle}\right)P_2^2+
\left(16 + \frac{- \lambda_1^2 m^{22}+ 2\lambda_1
\lambda_2 m^{12}-\lambda_2^2m^{11}}{3\triangle}\right)P_3^2\nonumber\\
& \rm + 12\left[   \left(4+ \frac{ \lambda_1 \lambda_2
m^{12}-\lambda_1^2 m^{22}}{3\triangle}\right)P_1+  \left(4
+\frac{\lambda_1 \lambda_{2}m^{12}-\lambda_2^2
m^{11}}{3\triangle}\right)P_2\right]
P_3 \nonumber\\
&\rm +24\left(3+\frac{\lambda_1\lambda_2m^{12}}{\triangle}\right)P_1
P_2- 24\left(V_1 e^{-\xi_1}+V_2
e^{-\xi_2}\right),\label{hamifrw_new}
\end{align}
 the parameter $\rm
\triangle$ is the same that was defined after equations
(\ref{mom_conjugados_26}). The form that the Hamiltonian density
(\ref{hamifrw_new}) acquires after applying the transformation
(\ref{trans_2}) into (\ref{hamifrw}) will, in the end, allows us to
obtain the solutions for this model. First, let's compute Hamilton's
equations, which read
\begin{align}
\rm \dot \xi_1=& \rm 24
\left(3-\frac{\lambda_1^2m^{22}}{\triangle}\right)P_1
+24\left(3+\frac{\lambda_1\lambda_2m_{12}}{\triangle}\right)P_2+ 12
\left(4+ \frac{\lambda_1 \lambda_2 m^{12}-\lambda_1^2
m^{22}}{3\triangle}\right)P_3,\nonumber\\
\rm \dot \xi_2=& \rm
24\left(3-\frac{\lambda_{2}^2\*m^{11}}{\triangle}\right)P_2
+24\left(3+\frac{\lambda_1\lambda_2m_{12}}{\triangle}\right)P_1+
12\left(4+\frac{\lambda_1\lambda_2
m^{12}-\lambda_2^2m^{11}}{3\triangle}\right)P_3, \nonumber\\
\rm \dot \xi_3 =&\rm 12\left[   \left(4+ \frac{ \lambda_1 \lambda_2
m^{12}-\lambda_1^2 m^{22}}{3\triangle}\right)P_1+  \left(4
+\frac{\lambda_1 \lambda_{2}m^{12}-\lambda_2^2
m^{11}}{3\triangle}\right)P_2\right] \nonumber\\
&+2 \left(16 + \frac{-
\lambda_1^2 m^{22}+ 2\lambda_1 \lambda_2
m^{12}-\lambda_2^2m^{11}}{3\triangle}\right)P_3,\nonumber\\
 \rm  \dot P_1=&\rm  -24 V_1e^{-\xi_1},\label{ecs_mov_2}\\
 \rm \dot P_2=&\rm -24V_2e^{-\xi_2},\nonumber \\
  \rm \dot P_3=&0,\nonumber
\end{align}
from this last set of equations is straightforward to see that $\rm
P_3$ is a constant. Taking the time derivative of the first equation
in (\ref{ecs_mov_2}), we obtain
\begin{equation}\rm
\rm \ddot \xi_1= \rm -576V_1 \left(3-\frac{\lambda_1^2
m^{22}}{\triangle}\right) e^{-\xi_1}
-576V_2\left(3+\frac{\lambda_1\lambda_2m^{12}}{\triangle}\right)
e^{-\xi_2}. \label{first}
\end{equation}
The main purpose of introducing the transformation (\ref{trans_2}) was to be able to separate the set of
equations arising from the Hamiltonian density (\ref{hamifrw_new}). To reach a solution for our problem
we set to zero the coefficient that is multiplying the mixed momenta term in (\ref{hamifrw_new}), which yield the
following constraint on the matrix element $\rm m^{12}$
\begin{equation}\label{m12}
\rm m^{12}=\frac{\lambda_1 \lambda_2}{6}\left(1 \pm \sqrt{1+
36 \frac{m^{11} m^{22}}{\lambda_1^2 \lambda_2^2}}\right),
\end{equation}
which implies that the second term in the square root of (\ref{m12}) is a real number, say $\ell=\rm 36(m^{11} m^{22}/\lambda_1^2 \lambda_2^2)$
 $\in\mathbb{R}^+$, giving the same weight to the matrix elements $\rm m^{11}$ and $\rm m^{22}$,
 whose values are $\rm m^{11}=(\sqrt{\ell}/6) \lambda_1^2$ and $\rm m^{22}=(\sqrt{\ell}/6) \lambda_2^2$. We are going
 to distinguish two possible scenarios for $\rm m^{12}$ as: $\rm m^{12}_+=(\lambda_1 \lambda_2/6)\left(1 + \sqrt{1+\ell}\right)>0$
 and $\rm m^{12}_-= - (\lambda_1 \lambda_2/6)\left(\sqrt{1+\ell}-1\right)<0$. This two choices of $\rm m^{12}$ enables us to have (what we called) a
 quintom like case and quintessence like case,
(however, the stiff matter scenario is dominant
 in all cases). With these two possible values for the matrix element $\rm m^{12}$ we can see that
$\triangle_+=-\frac{\lambda_1^2
\lambda_2^2}{18}\left(1+\sqrt{1+\ell}\right) <0$ for $\rm m^{12}_+$
and $\triangle_- =\frac{\lambda_1^2
\lambda_2^2}{18}\left(\sqrt{1+\ell}-1\right)>0$ for $\rm m^{12}_-$.


\subsection{Quintom like case}
\begin{figure}[ht!]
\begin{center}
\captionsetup{width=.9\textwidth}
\includegraphics[scale=0.375]{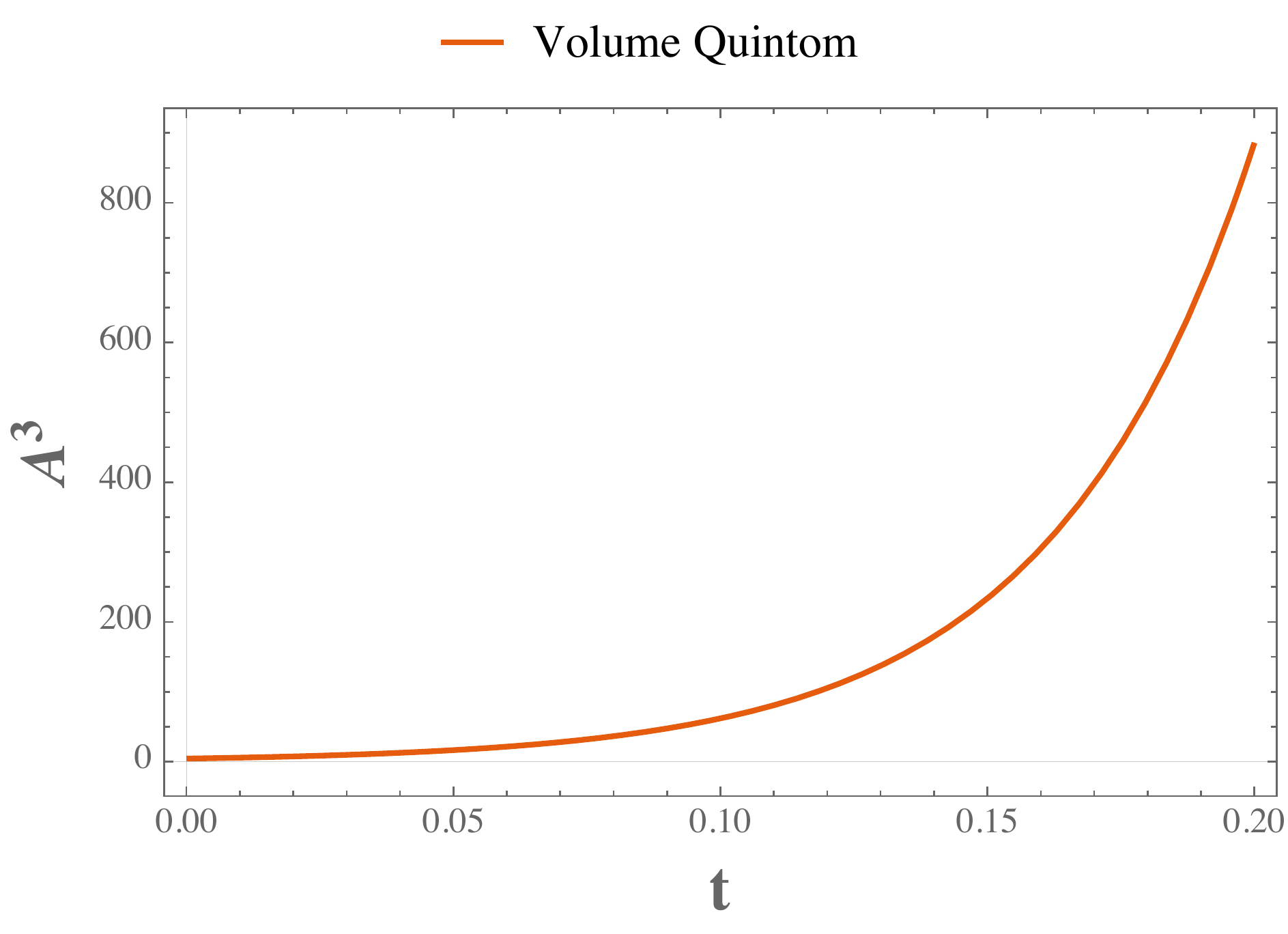}
\includegraphics[scale=0.4]{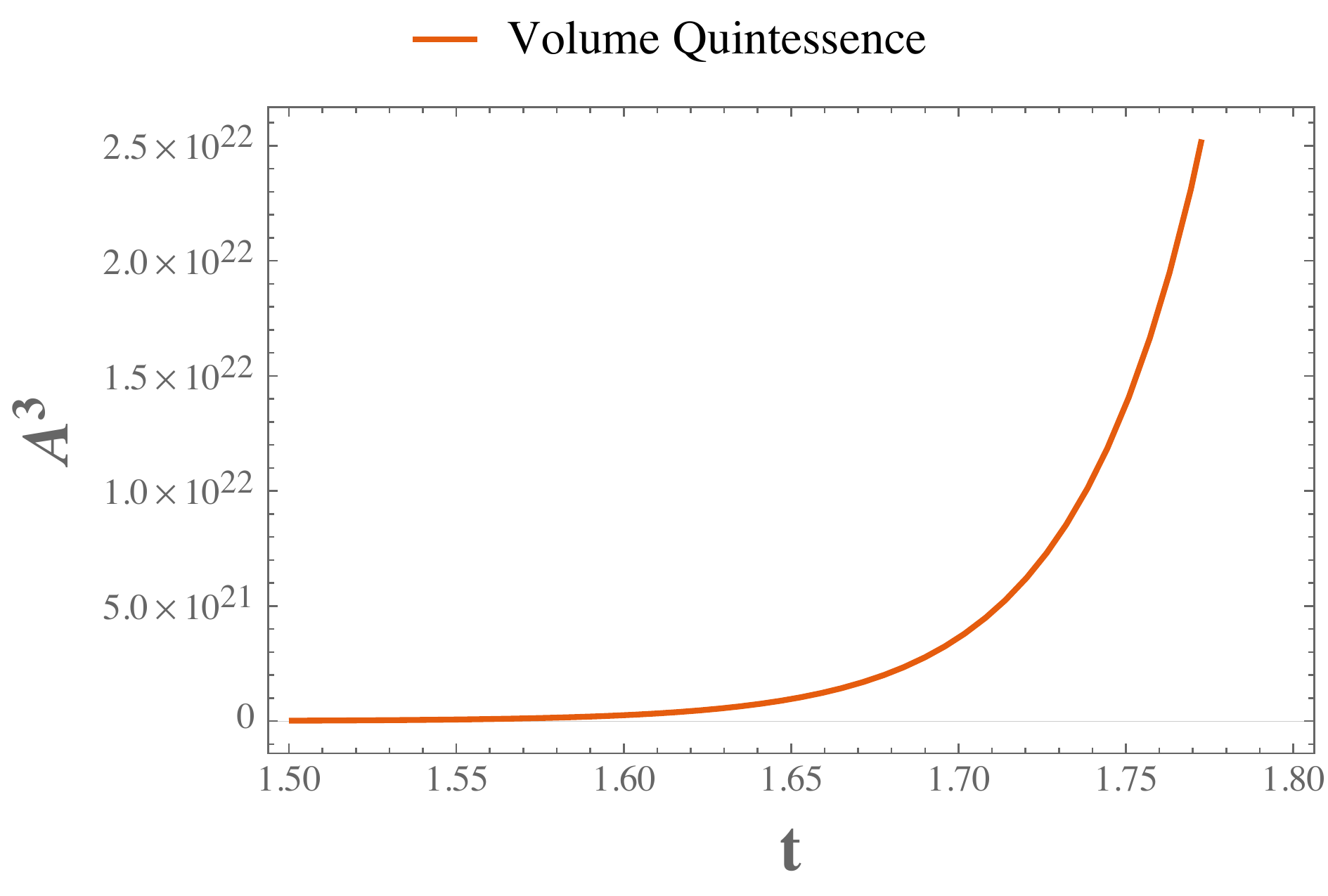}\\
\includegraphics[scale=0.4]{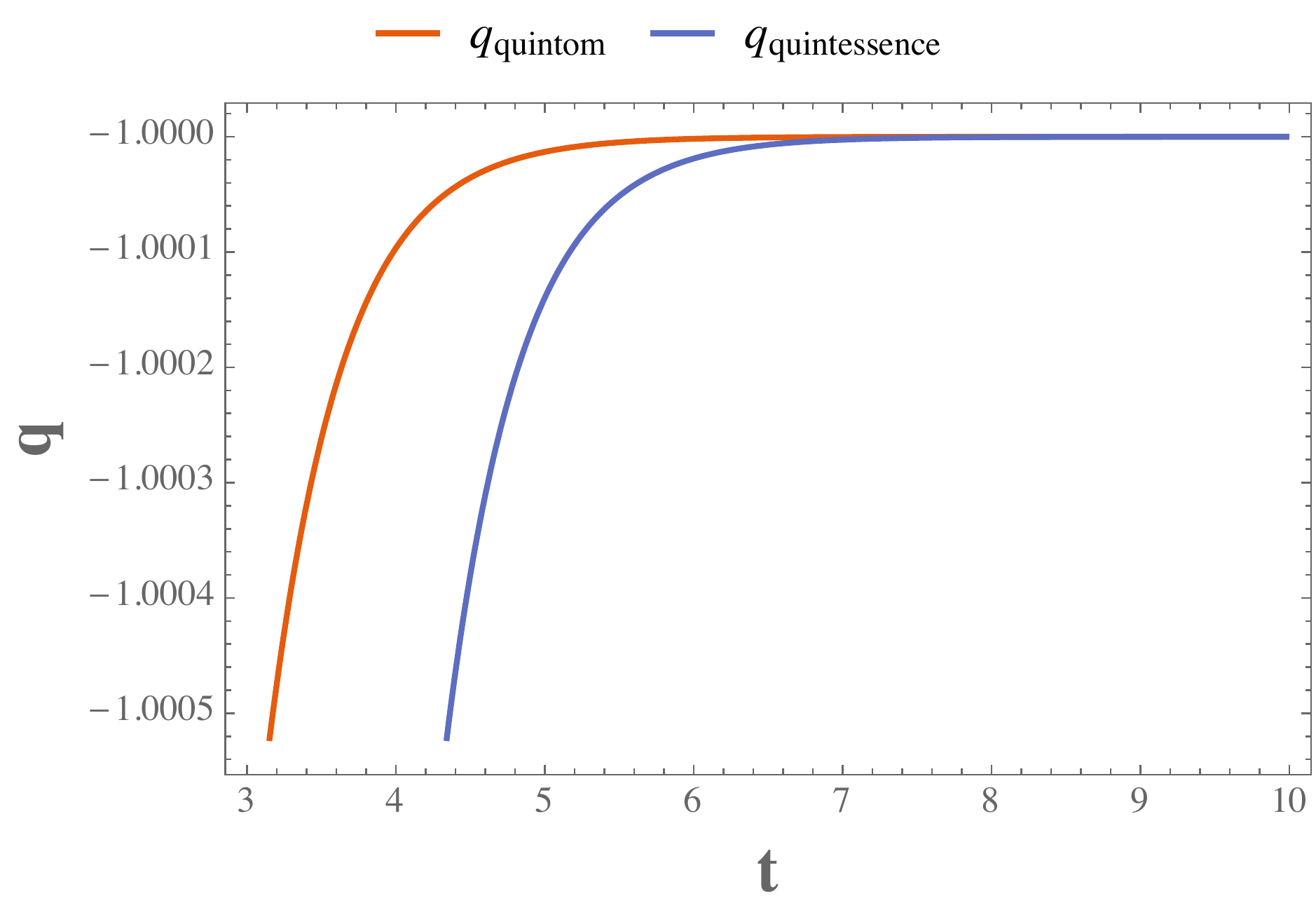}
\includegraphics[scale=0.39]{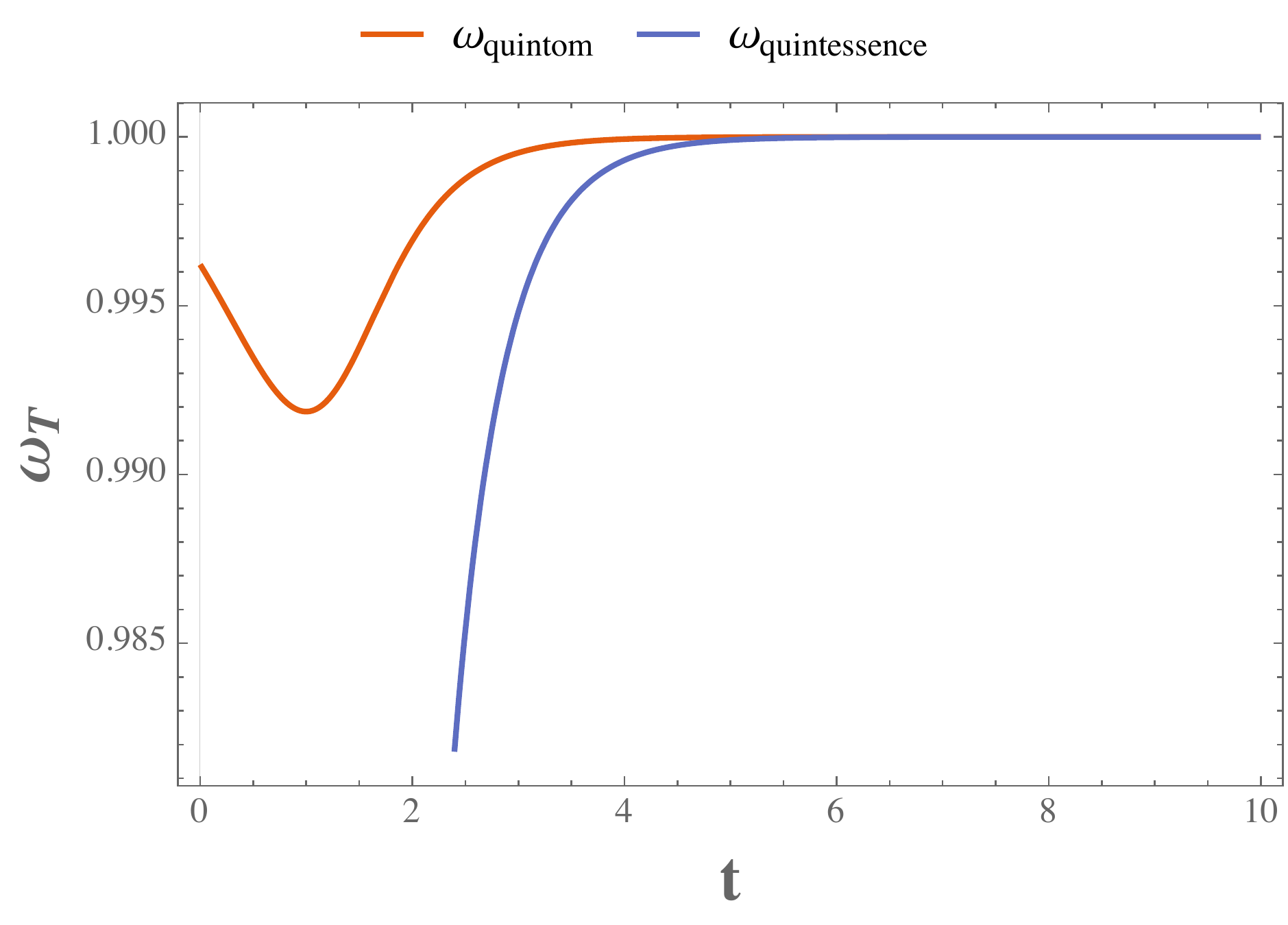}
\caption{In the top panel we can see the behavior the dynamical evolution of the volume for the quintom model
(left) and the quintessence model (right). In the bottom left panel its shown the deceleration parameter for
both models and in the bottom right panel we show the barotropic parameter for both models. The values taken for
this figures were $\ell=1$, $\rm p_3=1$, $\rm r_1=r_2=1$, $\rm q_1=q_2=1$, $\alpha_q=0.069$, $\beta_q=9.656$,
$\alpha_q^{\prime}=0.402$, $\beta_q^{\prime}=1.656$.}\label{vol_q_omega_ambos}
\end{center}
\end{figure}

We begin by analyzing the quintom like case, for which the matrix
element $\rm m^{12}_-= -b_\ell \lambda_1 \lambda_2$, and $\rm
b_\ell=(\sqrt{1+\ell}-1)/6$ has been defined in order to simplified
the calculations. Taking into account all the above, the Hamiltonian
density is rewritten as,
\begin{equation}   
\rm {\cal H}=  -\frac{P_1^2}{\mu_{_\ell}} -\frac{P_2^2}{\mu_{_\ell}}
+\left(48-\frac{1}{3c_{_\ell}} \right)\left(P_1+P_2 \right)
P_3+\left(16-\frac{1}{18c_{_\ell}}\right)P_3^2-24V_1e^{-\xi_1}-24V_2
e^{-\xi_2}, \label{hamifrwa}
\end{equation}
where we have define the parameters $\rm
\mu_{_\ell}=\frac{\sqrt{\ell}}{36\left[\left(1+\sqrt{1+\ell}\right)-\sqrt{\ell}\right]}$
and $\rm
c_{_\ell}=\frac{\sqrt{\ell}}{36\left[\left(1+\sqrt{1+\ell}\right)+\sqrt{\ell}\right]}$.
Thus, Hamilton equations for the new simplified coordinate $\rm
\xi_i$ are
\begin{eqnarray}\label{new_ecs_mov_2}
\rm \dot \xi_1&=& \rm -\frac{2P_1}{\mu_{_\ell}}
+\left(48-\frac{1}{3c_{_\ell}} \right)P_3, \nonumber\\
\rm \dot \xi_2&=& \rm -\frac{2P_2}{\mu_{_\ell}}
+\left(48-\frac{1}{3c_{_\ell}} \right)P_3, \\
\rm \dot \xi_3&=& \rm \left(48-\frac{1}{3c_{_\ell}}
\right)\left(P_1+P_2 \right) +
2\left(16-\frac{1}{18c_{_\ell}}\right)P_3, \nonumber
\end{eqnarray}
the equations for $\rm \dot P_1, \dot P_2$ and $\rm\dot P_3$ remain the same as in (\ref{ecs_mov_2}).
Taking the derivative of the first equation of (\ref{new_ecs_mov_2}) yields
\begin{equation}
\rm \ddot \xi_1= \frac{48V_1}{\mu_{_\ell}}  e^{-\xi_1},
\end{equation}
which has a solution of the form
\begin{equation}
\rm e^{-\xi_1}=\frac{\mu_{_\ell} r_1^2}{24 V_1} \, Sech^2\left(r_1
t-q_1\right). \label{solucion-xi1}
\end{equation}
From (\ref{new_ecs_mov_2}) we can see that $\dot\xi_2$ has the same functional structure as $\dot\xi_1$,
therefore its solution will be of the same form as (\ref{solucion-xi1}), so we have
\begin{equation}
\rm e^{-\xi_2}=\frac{\mu_{_\ell} r_2^2}{24 V_2} \, Sech^2\left(r_2
t-q_2\right), \label{solucion-xi2}
\end{equation}
where $\rm r_i$ and $\rm q_i$ (with $\rm i=1,2$) are integration constants, both at (\ref{solucion-xi1}) and (\ref{solucion-xi2}).
Reinserting these solutions into Hamilton equations for the momenta, we obtain
\begin{eqnarray}
\rm P_1 &=& \rm \alpha_1 - \mu_{_\ell} \,r_1 \, Tanh\left(r_1t-q_1\right), \label{solucion-p1}\\
\rm P_2 &=& \rm \alpha_2 - \mu_{_\ell}\,r_2 \, Tanh\left(r_2 t-q_2\right). \label{solucion-p2}
\end{eqnarray}
With (\ref{solucion-p1}) and (\ref{solucion-p2}), it can be easily check that the Hamiltonian is identically null when
\begin{equation}
\rm \alpha_1=\alpha_2=\frac{72\mu_{_\ell}-1}{6} p_3, \qquad
p_3^2=\frac{\mu_\ell(r_1^2+ r_2^2)}{4(72\mu_{_\ell}+1)}.
\end{equation}
Now we are in position write the solutions for the $\rm \xi_i$ coordinates, which read
\begin{align}
\rm \xi_1 &= \rm \beta_1 +Ln\left[ Cosh^2\left(r_1 t -q_1 \right)\right],  \label{xi1}\\
\xi_2 &= \rm \beta_2+Ln\left[ Cosh^2\left(r_2 t -q_2 \right)\right], \label{xi2}\\
\rm \xi_3&=\rm \beta_3 + p_3\left[16 \left(1+72\mu_{_\ell}
\right)-8\frac{\mu_{_\ell}}{c_{_{\ell}}}\right]\Delta t\\
&-\left(48-\frac{1}{3c_{_\ell}}\right)\mu_{_\ell}\,
\rm Ln\,\left[Cosh\left(r_1t-q_1 \right)\,Cosh\left(r_2 t-q_2 \right)
\right],
\end{align}
here the $\rm \beta_i$, with $\rm i=1,2,3$, terms are constants
coming from integration. Applying the inverse canonical
transformation we obtain the solutions in the original variables
$\rm (\Omega, \phi_1, \phi_2)$ as
\begin{equation}\label{sols_quintom}
\begin{split}
\rm \Omega &= \rm\Omega_0 +\frac{1}{12} Ln\left[ Cosh^2 \left(r_1 t
-q_1 \right) Cosh^2\left(r_2 t -q_2 \right)\right]
-\frac{1}{2}p_3\left[16 \left(1+72\mu_{_\ell}
\right)-8\frac{\mu_{_\ell}}{c_{_{\ell}}}\right]  \Delta t \\
& \rm + \frac{1}{2}\left(48 -\frac{1}{3c_{_\ell}}\right)\mu_\ell\,
Ln\,\left[Cosh\left(r_1t-q_1 \right)\,Cosh\left(r_2 t-q_2 \right) \right],\\
\rm \phi_1 &=\rm \phi_{10}+\frac{1}{2\lambda_1}\biggl[ Ln\left[Cosh^6\left(r_1t-q_1\right)Cosh^2\left(r_2 t -q_2\right)\right]
+6\mu_\ell \left(48-\frac{1}{3c_{_\ell}}\right) \times \\
&\phantom{{}={}} \rm Ln\,\left[Cosh
\left(r_1t-q_1\right)\,Cosh\left(r_2 t-q_2 \right) \right]
\biggr]\rm - \frac{3}{\lambda_1}p_3\left[16 \left(1+72\mu_{_\ell}
\right)-8\frac{\mu_{_\ell}}{c_{_{\ell}}}\right]\Delta t,\\
\rm \phi_2 &= \rm
\phi_{20}+\frac{1}{2\lambda_2}\biggl[Ln\left[Cosh^2\left(r_1t-q_1\right)
Cosh^6\left(r_2 t
-q_2\right)\right] + 6\mu_\ell\left (48-\frac{1}{3c_{_\ell}}\right) \times \\
&\phantom{{}={}} \rm Ln\,\left[Cosh \left(r_1t-q_1
\right)\,Cosh\left(r_2 t-q_2 \right) \right] \biggr]\rm  -
\frac{3}{\lambda_2}p_3\left[16 \left(1+72\mu_{_\ell}
\right)-8\frac{\mu_{_\ell}}{c_{_{\ell}}}\right]\Delta t,
\end{split}
\end{equation}
where $\Omega_0, \phi_{10}$ and $\phi_{20}$ are given in terms of the $\beta_i$ constants as
\begin{equation}
\Omega_0=\frac{\beta_1+\beta_2-6\beta_3}{12}, \quad
\phi_{10}=\frac{3\beta_1+\beta_2-6\beta_3}{2\lambda_1},\quad
\phi_{20}=\frac{\beta_1+3\beta_2-6\beta_3}{2\lambda_2}.\label{phi2}
\end{equation}
Thus, the scale factor is given by
\begin{equation}
\rm A(t)=A_0\, Cosh^{\frac{1}{6}+\alpha_q}\left(r_1\,t-q_1
\right)\,Cosh^{\frac{1}{6}+\alpha_q}\left(r_2\,t-q_2 \right)\, \,
e^{-\beta_q\,\Delta t}, \label{1-1}
\end{equation}
where  $\rm
\beta_q=\frac{p_3}{2}\left(16(1+72\mu_\ell)-8\frac{\mu_\ell}{c_\ell}\right)
$ and
$\rm\alpha_q=\frac{1}{2}\left(48-\frac{1}{3c_{_\ell}}\right)\mu_\ell$.
The dynamical behavior of the volume function is presented in
Fig.~(\ref{vol_q_omega_ambos}). The deceleration parameter for this
quintom cosmological model becomes
\begin{equation}\label{q-quintom}
\rm q_{quintom}=-1 -\alpha_0\,\frac{\left(r_1^2\,Cosh^2(r_2\,t-q_2)+r_2^2\,Cosh^2(r_1\,t -q_1)\right)}{T},
\end{equation}
with
\begin{align}
\rm T&=& \rm \alpha_0^2\left(r_1^2 Sinh^2(r_1\,t-q_1)\,Cosh^2(r_2\,t-q_2)+r_2^2\,Sinh^2(r_2\,t-q_2)\,Cosh^2(r_1\,t-q_1)\right)\nonumber\\
&&\rm + 2\alpha_0\, Sinh(r_1\,t-q_1)\,Sinh(r_2\,t-q_2)\left\{\alpha_0r_1r_2 Cosh(r_1t-q_1)\,Cosh(r_2t-q_2)+ \right. \nonumber\\
&&\left. \rm  \beta_q \left[r_2\,Cosh(r_2 t-q_2)\,Sinh(r_1t-q_1)+r_1Sinh(r_2t-q_2)\,Cosh(r_1t-q_1)\right]\right\}\nonumber\\
&&\rm +\beta_q^2\,Cosh^2(r_1\,t-q_1)\,Cosh^2(r_2\,t-q_2),
\end{align}
and $\rm\alpha_0=\frac{1}{6}+\alpha_q$. For this model, the barotropic parameter $\omega_T$ takes the form
\begin{equation}
\rm \omega_{quintom}=1 -\frac{2}{3}\alpha_0\,\frac{\left(r_1^2\,Cosh^2(r_2\,t
-q_2)+r_2^2\,Cosh^2(r_1\,t -q_1)\right)}{T}.
\end{equation}
The corresponding behavior of the deceleration and the barotropic
parameters are also shown in the figure (\ref{vol_q_omega_ambos})
below.


\subsection{Quintessence like case }

Now we turn our attention to the quintessence like case, for which
the matrix element $\rm{m^{12}_+= q_\ell \lambda_1 \lambda_2}$, and
$\rm{q_\ell=\left(1 + \sqrt{1+\ell}\right)/6}$ has been defined to
make the calculations simpler. The Hamiltonian density describing
this quintessence model is rewritten as
\begin{equation}
\rm {\cal H}=  \frac{P_1^2}{\nu_{_\ell}} + \frac{P_2^2}{\nu_{_\ell}}
+\left(48-\frac{1}{3c_{_\ell}^{\prime}} \right)\left(P_1+P_2 \right)
P_3+\left(16-\frac{1}{18c_{\ell}^{\prime}}\right)P_3^2-24V_1e^{-\xi_1}-24V_2
e^{-\xi_2}, \label{hamifrwb}
\end{equation}
here we define the parameters
$\rm\nu_{\ell}=\frac{\sqrt{\ell}}{36\left(\sqrt{1+\ell}+\sqrt{\ell}-1\right)}$
and $\rm
c_{_\ell}^{\prime}=\frac{\sqrt{\ell}}{36\left(+\sqrt{\ell}+1-\sqrt{1+\ell}\right)}$.

From (\ref{hamifrwb}) we can calculate Hamilton equations for the
phase space spanned by $\rm (\xi_i, P_i)$, given by
\begin{eqnarray}
\rm \dot \xi_1&=& \rm \frac{2P_1}{\nu_{_\ell}}
+\left(48-\frac{1}{3c_{_\ell}^{\prime}} \right)P_3, \nonumber \\
\rm \dot \xi_2&=& \rm \frac{2P_2}{\nu_{_\ell}}
+\left(48-\frac{1}{3c_{_\ell}^{\prime}} \right)P_3, \label{xi_quintessence}\\
\rm \dot \xi_3&=& \rm \left(48-\frac{1}{3c_{_\ell}^{\prime}}
\right)\left(P_1+P_2 \right) +
2\left(16-\frac{1}{18c_{_\ell}^{\prime}}\right)P_3,\nonumber
\end{eqnarray}
as in the quintom case $\rm \dot P_1, \dot P_2$ and $\rm \dot P_3$ remain the same as in (\ref{ecs_mov_2}).
Proceeding in a similar way as in the previous case, we take the derivative of the first equation in (\ref{xi_quintessence}), obtaining
\begin{equation}
\rm \ddot \xi_1= -\frac{48V_1}{\nu_{_\ell}}  e^{-\xi_1},
\end{equation}
which the corresponding solution is
\begin{equation}
\rm e^{-\xi_1}=\frac{\nu_{_\ell} r_1^2}{24 V_1} \,
Csch^2\left(r_1t-q_1\right). \label{solution-xi1-1}
\end{equation}
Also in this quintessence like setting, the functional
form of $\dot\xi_2$ is the same as $\dot\xi_1$, indicating that the solution is of
the same type as (\ref{solution-xi1-1}), that is
\begin{equation}
\rm e^{-\xi_2}=\frac{\nu_{_\ell} r_2^2}{24 V_2} \, Csch^2\left(r_2
t-q_2\right), \label{solution-xi2-1}
\end{equation}
in (\ref{solution-xi1-1}) and (\ref{solution-xi2-1}) the $\rm r_i$
and $\rm q_i$ (with $\rm i=1,2$) are constants coming from integration. With (\ref{solution-xi1-1})
and (\ref{solution-xi2-1}) at hand, we can reinsert them into
Hamilton equations for the momenta, giving
\begin{eqnarray}
\rm P_1 &=& \rm -a_1 + \nu_{_\ell} \,r_1 \, Coth\left(r_1 t-q_1
\right), \label{solution-p1-1} \\
\rm P_2 &=& \rm -a_2 + \nu_{_\ell}\,r_2 \, Coth\left(r_2 t-q_2
\right), \label{solution-p2-1}
\end{eqnarray}
where it can be easily verify that with (\ref{solution-p1-1}) and (\ref{solution-p2-1}) at hand the Hamiltonian is identically zero when
\begin{equation}
\rm a_1=a_2=\frac{72\nu_{\ell}+1}{6} p_3, \qquad
p_3^2=\frac{\nu_{_\ell}(r_1^2+ r_2^2)}{4(72\nu_{_\ell}-1)}.
\end{equation}
So, the solutions for the $\rm \xi_i$ coordinates become
\begin{align}
\rm \xi_1 &= \rm \beta_1 +Ln\left[ Sinh^2
\left(r_1 t -q_1 \right)\right],  \label{xi1-1}\\
\xi_2 &= \rm \beta_2+Ln\left[ Sinh^2
\left(r_2 t -q_2 \right)\right], \label{xi2-1}\\
\rm \xi_3&= \rm \beta_3 -p_3
\left[16\left(72\nu_{\ell}-1\right)-8\frac{\nu_{\ell}}{c_{\ell}^{\prime}}\right]
\Delta t\\
&+ \left(48 -\frac{1}{3c_{\ell}^{\prime}}\right)\,\nu_{\ell}
\rm Ln\,\left[Sinh \left(r_1t-q_1 \right)\,Sinh\left(r_2 t-q_2 \right)
\right],
\end{align}
where $\rm \beta_i$ are integration constants. After applying the inverse canonical transformation we
get the solutions in terms of the original variables $\rm (\Omega, \phi_1,
\phi_2)$ as
\begin{equation}\label{sols_quintessence}
\begin{split}
\rm \Omega &= \rm  \Omega_0 +\frac{1}{12} Ln\left[ Sinh^2 \left(r_1t -q_1 \right) Sinh^2\left(r_2 t -q_2 \right)\right]
+\frac{1}{2}p_3\left[16\left(72\nu_{\ell}-1\right)-8\frac{\nu_{\ell}}{c_{\ell}^{\prime}}\right]\Delta t \\
&\rm - \frac{1}{2}\left(48 -\frac{1}{3c_{\ell}^{\prime}}\right)\,\nu_{\ell}
Ln\,\left[Sinh \left(r_1t-q_1 \right)\,Sinh\left(r_2 t-q_2 \right) \right], \\
 \rm \phi_1 &= \rm \phi_{10}+\frac{1}{2\lambda_1}\biggl[ Ln\left[Sinh^6\left(r_1t-q_1\right)
Sinh^2\left(r_2 t -q_2\right)\right] -
6\left(48-\frac{1}{3c_{\ell}^{\prime}}\right)\nu_{\ell}\times \\
&\phantom{{}={}} \rm Ln\,\left[Sinh \left(r_1t-q_1\right)\,Sinh\left(r_2 t-q_2 \right) \right] \biggr]\rm
+ \frac{3}{\lambda_1}p_3 \left[16\left(72\nu_{\ell}-1\right)-8\frac{\nu_{\ell}}{c_{\ell}^{\prime}}\right]\Delta t, \\
\rm \phi_2 &= \rm
\phi_{20}+\frac{1}{2\lambda_2}\biggl[Ln\left[Sinh^2\left(r_1t-q_1\right)
Sinh^6\left(r_2 t
-q_2\right)\right] - 6\left(48-\frac{1}{3c_{\ell}^{\prime}}\right)\nu_{\ell} \times \\
&\phantom{{}={}} \rm Ln\,\left[Sinh
\left(r_1t-q_1\right)\,Sinh\left(r_2 t-q_2 \right) \right]
\biggr]\rm + \frac{3}{\lambda_2}p_3
\left[16\left(72\nu_{\ell}-1\right)-8\frac{\nu_{\ell}}{c_{\ell}^{\prime}}\right]\Delta
t,
\end{split}
\end{equation}
where $\Omega_0, \phi_{10}$ and $\phi_{20}$ are given in terms of the $\beta_i$ constants as
\begin{equation}
\Omega_0=\frac{\beta_1+\beta_2-6\beta_3}{12},\quad
 \phi_{10}=\frac{3\beta_1+\beta_2-6\beta_3}{2\lambda_1},\quad
 \phi_{20}=\frac{\beta_1+3\beta_2-6\beta_3}{2\lambda_2}.\label{phi2-11}
\end{equation}
In this case, the scale factor acquires the form
\begin{equation}
\rm A(t)=A_0\, Sinh^{\frac{1}{6}-\alpha_q^{\prime}}\left(r_1\,t-q_1
\right)\,Sinh^{\frac{1}{6}-\alpha_q^{\prime}}\left(r_2\,t-q_2 \right)\, \,
e^{\beta_q^{\prime}\,\Delta t}, \label{1-2}
\end{equation}
where  $\rm
\beta_q^{\prime}=\frac{p_3}{2}\left(16(72\nu_{_\ell}-1)-8\frac{\nu_{_\ell}}{c_{_\ell}^{\prime}}
\right) $ and  $\rm \alpha_q^{\prime}=\frac{1}{2}\left(48
-\frac{1}{3c_{_\ell}^{\prime}}\right)\nu_{_\ell}$.\\\\
The deceleration parameter for this case is
\begin{equation} \rm q_{quintessence}=-1 -\frac{\alpha_1^{\prime}\left(r_1^2\,Sinh^2(r_2\,t
-q_2)+r_2^2\,Sinh^2(r_1\,t -q_1)\right)}{Q},
\end{equation}
with
\begin{align}
\rm Q&=& \rm \alpha_1^{\prime 2}\left(r_1^2
Sinh^2(r_2\,t-q_2)\,Cosh^2(r_1\,t-q_1)+r_2^2\,Sinh^2(r_1\,t-q_1)\,Cosh^2(r_2\,t-q_2)\right)\nonumber\\
&&\rm + 2\alpha_1^{\prime}\, Sinh(r_1\,t-q_1)\,Sinh(r_2\,t-q_2)\left\{
+\alpha_1^{\prime}
r_1r_2 Cosh(r_1t-q_1)\,Cosh(r_2t-q_2)+ \right. \nonumber\\
&&\left. \rm  \beta_q^{\prime} \left[r_2\,Cosh(r_2 t-q_2)\,Sinh(r_1t-q_1)+r_1
Sinh(r_2t-q_2)\,Cosh(r_1t-q_1)
\right]\right\}\nonumber\\
&&\rm +\beta_q^{\prime 2}\,
Sinh^2(r_1\,t-q_1)\,Sinh^2(r_2\,t-q_2).
\end{align}
where the constant  $\rm \alpha_1^{\prime}=\alpha_q^{\prime}-\frac{1}{6}.$
We are able to calculate the barotropic parameter employing the equation (\ref{omegat}), giving
\begin{equation}
\rm \omega_{quintessence}=1
-\frac{2}{3}\frac{\alpha_1^{\prime}\left(r_1^2\,Sinh^2(r_2\,t
-q_2)+r_2^2\,Sinh^2(r_1\,t -q_1)\right)}{Q},
\end{equation}
in the lower right panel of the figure shown below we can see the
behavior of the barotropic parameter.

It is clear that the standard quintessence model with two scalar
fields cannot be reproduce under this approach, because when we set
$\rm{m^{12}=0}$, this imply that parameter $\ell$ is equal to
zero, then, the matrix elements ${\rm m^{11}=m^{22}}$ are
also zero, this was the challenge to resolve.

\section{Quantum Approach \label{qsolutions}}
In this section we present the quantum version of the classical
cosmological models studied above along with its solutions. Since we
already have the classical Hamiltonian density, the quantum
counterpart can be obtained making the usual replacement $\rm
\Pi_{q^\mu}=-i\hbar \partial_{q^\mu}$. First we modified the
classical Hamiltonian density (\ref{first-hamifrw}) in order to
consider the factor ordering problem between the function $\rm
e^{-3\Omega}$ and its moment $\rm \Pi_\Omega$, introducing the
linear term as $\rm  e^{-3\Omega}\Pi_\Omega^2 \to
e^{-3\Omega}\left[\Pi_\Omega^2 +Qi\hbar \Pi_\Omega \right]$ where Q
is a real number that measure the ambiguity in the factor ordering.

\subsection{Quantum Quintessence-K-essence standard case}\label{quantum_QK}
\begin{figure}[ht!]
\begin{center}
\captionsetup{width=.8\textwidth}
\includegraphics[scale=0.8]{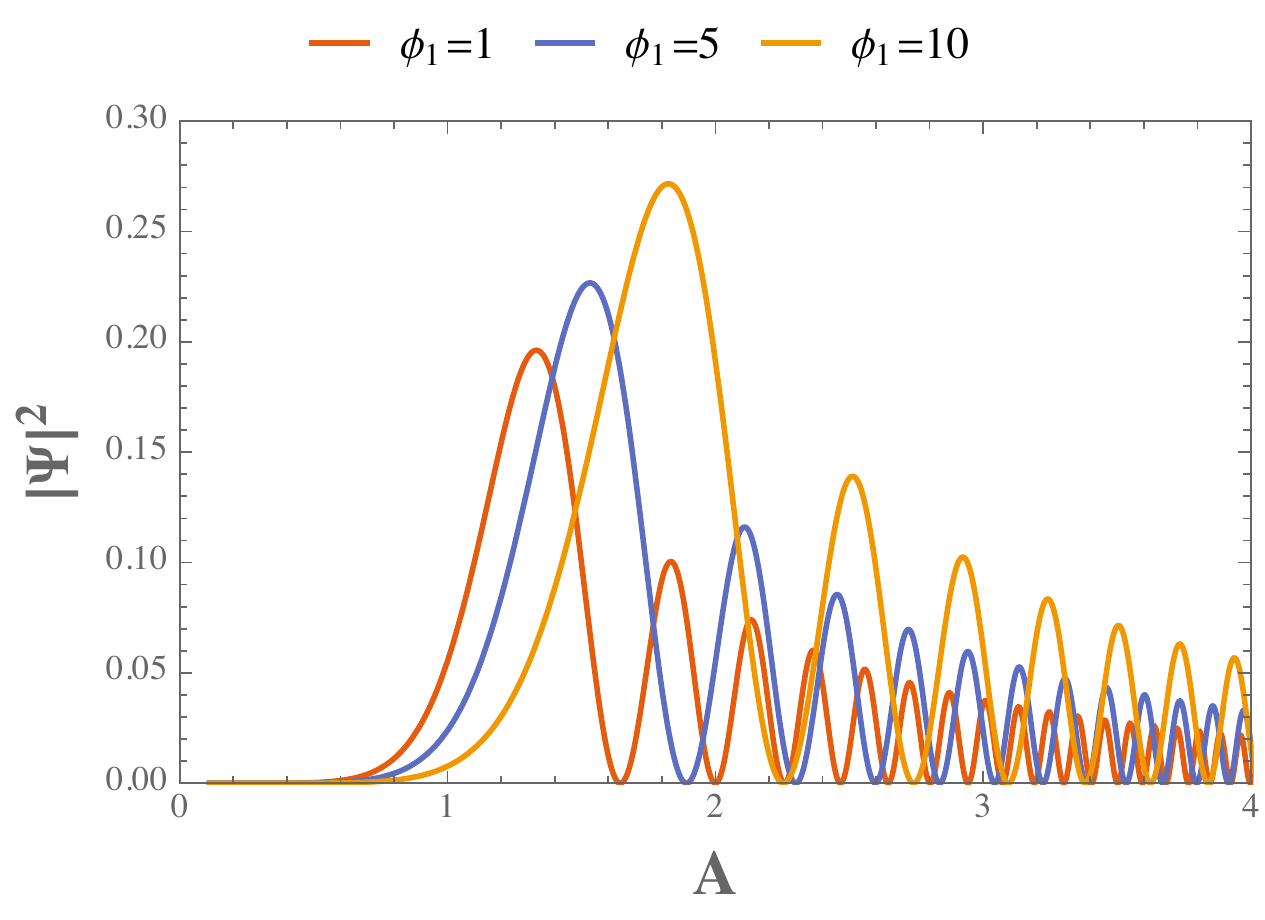}
\caption{Behaviour of density probability considering some values in
the scalar field $\phi_1=1,5, 10$.}\label{cortes}
\end{center}
\end{figure}

The quantum version for the first cosmological model  we employ the
modified Hamiltonian density,
\begin{equation}
\rm {\cal H}=  \Pi_\Omega^2 +Qi\hbar \Pi_\Omega-12
 \Pi_{\phi_1}^2 - 12
\Pi_{\phi_2}^2 -24V_1 e^{6\Omega-\lambda_1\phi_1}  \,,
\label{mod-hamifrw}
\end{equation}
at this point, in order to obtain the Wheeler-DeWitt equation, we implement the following change of variables
$\rm(\Omega,\phi_1,\phi_2)\leftrightarrow (\xi_1,\xi_2,\xi_3)$
\begin{equation}\label{uuno}
\begin{split}
\rm \xi_1 &=\rm6\Omega-\lambda_1 \phi_1, \\
\xi_2&= \Omega, \\
\xi_3&=\phi_2,\\
\end{split}
\qquad\longleftrightarrow\qquad
\begin{split}
\rm \Omega &=\rm \xi_2, \\
\phi_1&=\frac{-\xi_1+6\xi_2}{\lambda_1}, \\
\phi_2&= \xi_3,
\end{split}
\end{equation}
and also, obtaining a new set of conjugate momenta (in the same
manner as (\ref{momenta-s})), of the variables $\rm
(\xi_1,\xi_2,\xi_3)$, namely $\rm (P_1,P_2,P_3)$, which read
\begin{equation}
\rm \Pi_\Omega =   6 P_1 + P_2, \qquad  \Pi_{\phi_1} = -\lambda_1
P_1, \qquad  \Pi_{\phi_2} = P_3, \label{new-momenta}
\end{equation}
which in turn transform the Hamiltonian density (\ref{mod-hamifrw})
as
\begin{equation}
\rm {\cal H} = \rm  12 \left(3-\lambda_1^2 \right)P_1^2+ P_2^2 +12
P_1 P_2-12 P_3^2+i\hbar Q(6p_1 + p_2)- 24V_1 e^{\xi_1}.
\label{new-hami}
\end{equation}
Introducing the replacement $\rm \Pi_{q^\mu}=-i\hbar\partial_{q^\mu}$, the WDW equation becomes
\begin{equation}
\begin{split}
\rm {\cal H}\Psi = -12 \hbar^2\left(3-\lambda_1^2
\right)\frac{\partial^2\Psi}{\partial \xi_1^2}-\hbar^2
\frac{\partial^2\Psi}{\partial \xi_2^2} - 12\hbar^2
\frac{\partial^2\Psi}{\partial \xi_1 \partial \xi_2}+\\12 \hbar^2
\frac{\partial^2\Psi}{\partial \xi_3^2}+Q\hbar^2
\left(6\frac{\partial \Psi}{\partial \xi_1}+\frac{\partial
\Psi}{\xi_2}\right)- 24V_1 e^{\xi_1}\Psi=0, \label{q-mod}
\end{split}
\end{equation}
due that the scalar potential does not depend on the coordinates
$\rm (\xi_2,\xi_3)$, we propose the following ansatz for the wave
function $\rm \Psi(\xi_1,\xi_2,\xi_3)=e^{-(a_2\xi_2+a_3\xi_3)/\hbar}
G(\xi_1)$ where $\rm a_2$ and $\rm a_3$ are arbitrary constants.
Introducing the mentioned ansatz in (\ref{q-mod}) we have that
\begin{equation}
\rm -12 \hbar^2\left(3-\lambda_1^2 \right)\frac{1}{G}\frac{d^2G}{d
\xi_1^2} +6  \hbar\left(2 a_2+\hbar Q\right) \frac{1}{G}
\frac{dG}{d\xi_1}-a_2(a_2+\hbar Q)+12 a_3^2- 24V_1 e^{\xi_1}=0,
\nonumber
\end{equation}
where we also divided the whole equation by the ansatz; this in turn
leads us to the following differential equation
\begin{equation}
\rm\frac{d^2G}{d\xi_1^2} -\frac{2a_2+ \hbar
Q}{2\hbar(3-\lambda_1^2)}\frac{dG}{d\xi_1}
+\frac{1}{12\hbar^2(3-\lambda_1^2)} \left[24V_1e^{\xi_1}+ \eta
\right]G=0,\label{qauntum_psi_1}
\end{equation}
here $\eta=a_2(a_2+\hbar Q)-12 a_3^2$. The last equation can be casted as $\rm y^{\prime \prime} + a y^\prime + \left(b e^{\kappa x } +c
\right)y=0$ (and whose solutions will depend on the value of $\lambda_1$) \cite{polyanin}, where
\begin{equation}
\rm y=Exp\left({-\frac{ax}{2}}\right) Z_\nu \left(\frac{2\sqrt{b}}{\kappa} e^{\frac{\kappa x}{2}} \right),
\end{equation}
here $\rm Z_\nu$ is the Bessel function and $\nu=\sqrt{a^2-4c}/\kappa$ being the order.
The corresponding relations between the coefficients of (\ref{qauntum_psi_1}) and $\rm a,b,c$ and $\kappa$ are
\begin{align}
\rm a &=\rm \left\{
\begin{tabular}{ll}
$\rm \frac{2a_2+ \hbar Q}{2\hbar(\lambda_1^2-3)},$ &when $\lambda_1^2 > 3$\nonumber \\ \\
$\rm  -\frac{2a_2+ \hbar Q}{2\hbar(3-\lambda_1^2)},$ & when $\lambda_1^2 < 3$
\end{tabular}
\right. \\  \\
\rm b &=\rm \left\{
\begin{tabular}{ll}
$\rm -\frac{2V_0}{\hbar^2(\lambda^2-3)},$ & when \,\,$\lambda_1^2 > 3$\\ \\
$\rm \frac{2V_0}{\hbar^2(3-\lambda^2)},$ & when \,\,$\lambda_1^2 <3$ \nonumber
\end{tabular}
\right. \\  \\
\rm c &= \rm \left\{
\begin{tabular}{ll}
$\rm -\frac{\eta}{12 \hbar^2\left(\lambda_1^2-3\right)}$, & when \,\,$\lambda_1^2 > 3$\\ \\
$\rm \frac{\eta}{12\hbar^2\left(3-\lambda_1^2\right)}$,& when \,\,$\lambda_1^2 < 3$ \nonumber
\end{tabular}
\right.\\ \\
\kappa&=1,
\end{align}
according to the constant b, the solution to the function $\rm G$ becomes
\begin{align}
\rm G(\xi_1) &=\rm Exp\left(-\frac{2a_2+ \hbar
Q}{4\hbar(\lambda_1^2-3)}\xi_1 \right)\,\,\, K_{\nu_1}\left(
\frac{2}{\hbar}\sqrt{\frac{2V_0}{ \lambda^2-3 }}\,\, e^{\frac{\xi_1}{2}} \right), \qquad \lambda_1^2 > 3 \label{k0}\\
\rm G(\xi_1) &=\rm Exp\left(\frac{2a_2+ \hbar
Q}{4\hbar(3-\lambda_1^2)\xi_1} \right)\,\,\,
J_{\nu_2}\left(\frac{2}{\hbar}\sqrt{\frac{2V_0}{ 3-\lambda_1^2}}\,\,
 e^{\frac{\xi_1}{2}} \right), \qquad \lambda_1^2 < 3 \label{j0}
\end{align}
and the wavefunction takes the form
\begin{align}
\rm \Psi_{\nu_1} &=\rm  Exp\left(-\frac{2a_2+ \hbar
Q}{4\hbar(\lambda_1^2-3)}\xi_1 - \frac{a_2 \xi_2 + a_3 \xi_3}{\hbar}
\right)\,\,\, K_{\nu_1}\left(
\frac{2}{\hbar}\sqrt{\frac{2V_0}{ \lambda_1^2-3 }}\,\, e^{\frac{\xi_1}{2}} \right), \quad \lambda_1^2 > 3 \label{k1}\\
\rm \Psi_{\nu_2} &=\rm  Exp\left(\frac{2a_2+ \hbar
Q}{4\hbar(3-\lambda_1^2)}\xi_1 - \frac{a_2 \xi_2 + a_3 \xi_3}{\hbar}
\right)\,\,\, J_{\nu_2}\left(\frac{2}{\hbar}\sqrt{\frac{2V_0}{
3-\lambda^2}}\,\,
 e^{\frac{\xi_1}{2}} \right), \quad \lambda_1^2 < 3. \label{j1}
\end{align}
where $\rm \nu_1=\sqrt{\left(-\frac{2a_2+ \hbar
Q}{4\hbar(\lambda_1^2-3)} \right)^2 +
\frac{4\eta}{12\hbar^2(\lambda_1^2-3)}}$ and
 $\rm \nu_2=\sqrt{\left(\frac{2a_2+ \hbar
Q}{4\hbar(3-\lambda_1^2)} \right)^2 -
\frac{4\eta}{12\hbar^2(3-\lambda_1^2)}}$ are the corresponding orders of the wave function.
Applying the inverse transformation on the variables $\rm \xi_i$, we can write the wave function in terms
of the original variables $(\rm A=e^\Omega,\phi_i)$, which read
\begin{align}
\rm \Psi_{\nu_1} =\rm  A^{-\alpha_1 }\,Exp\left(\frac{2a_2+ \hbar
Q}{4\hbar(\lambda_1^2-3)}\lambda_1 \phi_1 - \frac{ a_3}{\hbar}\phi_2
\right)\,\,\, K_{\nu_1}\left(
\frac{2}{\hbar}\sqrt{\frac{2V_0}{ \lambda_1^2-3 }}\,\, A^{3}e^{\frac{\lambda_1}{2}\phi_1} \right), \qquad \lambda_1^2 > 3 \label{k2}\\
\rm \Psi_{\nu_2} =\rm  A^{-\alpha_2 }\, Exp\left(-\frac{2a_2+
\hbar Q}{4\hbar(\lambda_1^2-3)}\lambda_1 \phi_1 - \frac{
a_3}{\hbar}\phi_2 \right)\,\,\,
J_{\nu_2}\left(\frac{2}{\hbar}\sqrt{\frac{2V_0}{ 3-\lambda^2}}\,\,
 A^{3}e^{\frac{\lambda_1}{2}\phi_1} \right), \qquad \lambda_1^2 < 3. \label{j2}
\end{align}
with $\alpha_1=\frac{1}{\hbar}\left( a_2+\frac{3}{2}\frac{2a_2+\hbar
Q}{\lambda_1^2-3}\right)$ and $\alpha_2=\frac{1}{\hbar}\left(
a_2-\frac{3}{2}\frac{2a_2+\hbar Q}{3-\lambda_1^2}\right)$. The
behavior of the wave function of this model, when $\lambda_1^2<{3}$,
can be seen in the following figures. In Fig.~(\ref{density1}) it can
be observed that the probability density has a damped behavior,
which is a good characteristic in a wave function and this kind of
demeanor has been reported in \cite{soco2,sor}.
\begin{figure}[ht!]
\begin{center}
\captionsetup{width=.8\textwidth}
\includegraphics[scale=0.4]{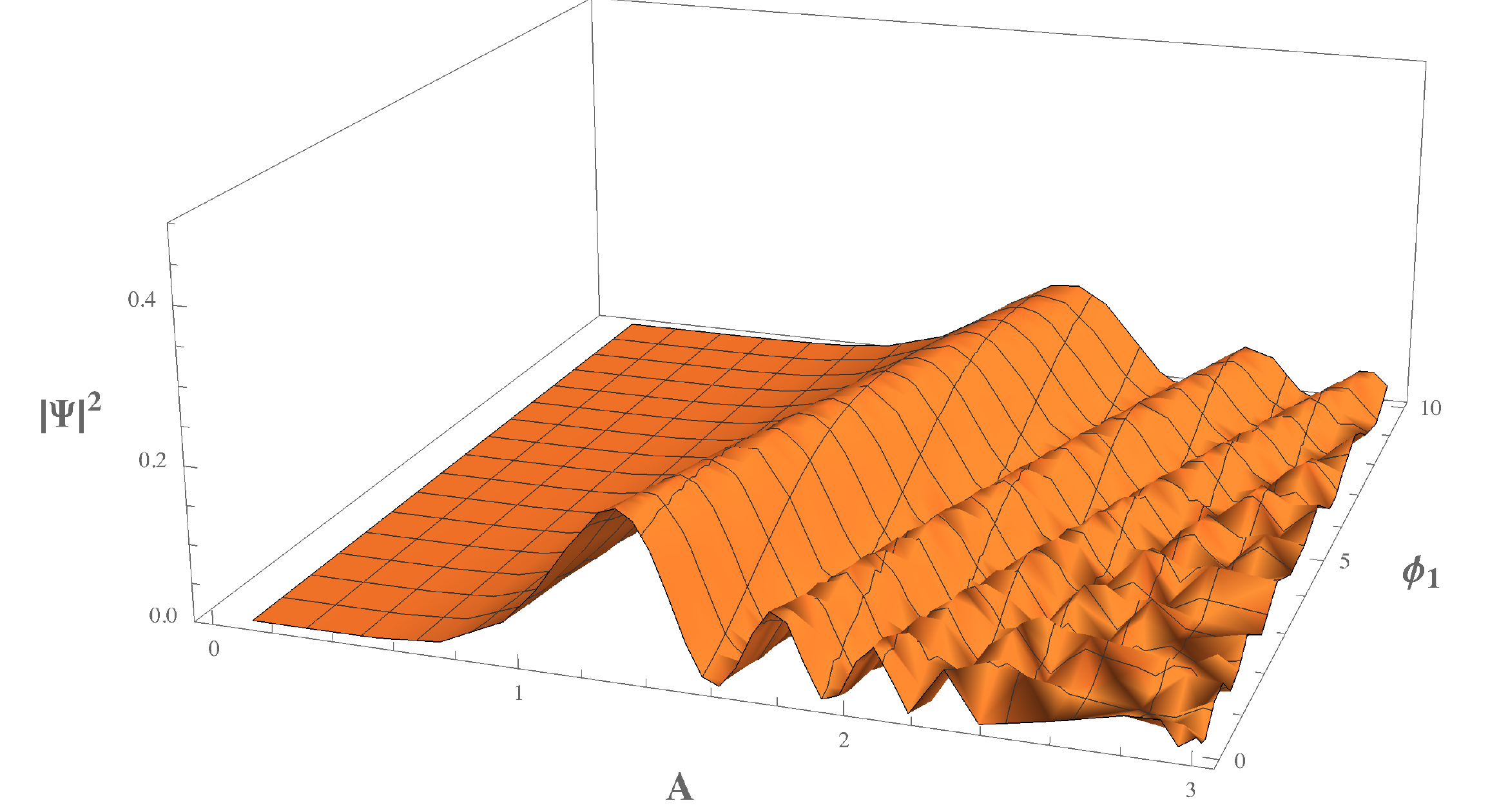}
\caption{Behaviour of density probability, with $\rm
Q=1,\,\lambda_1=0.21$, $\rm a_2=0.6,\, a_3=1$, $\rm \nu_2$.}
\label{density1}
\end{center}
\end{figure}
In Figure (\ref{cortes}) a 2D view of the probability density for
different values of $\phi_1$ its shown, we can also see the
importance of the existence scalar fields during primordial
inflation.

\begin{figure}[ht!]
\begin{center}
\captionsetup{width=.8\textwidth}
\includegraphics[scale=0.35]{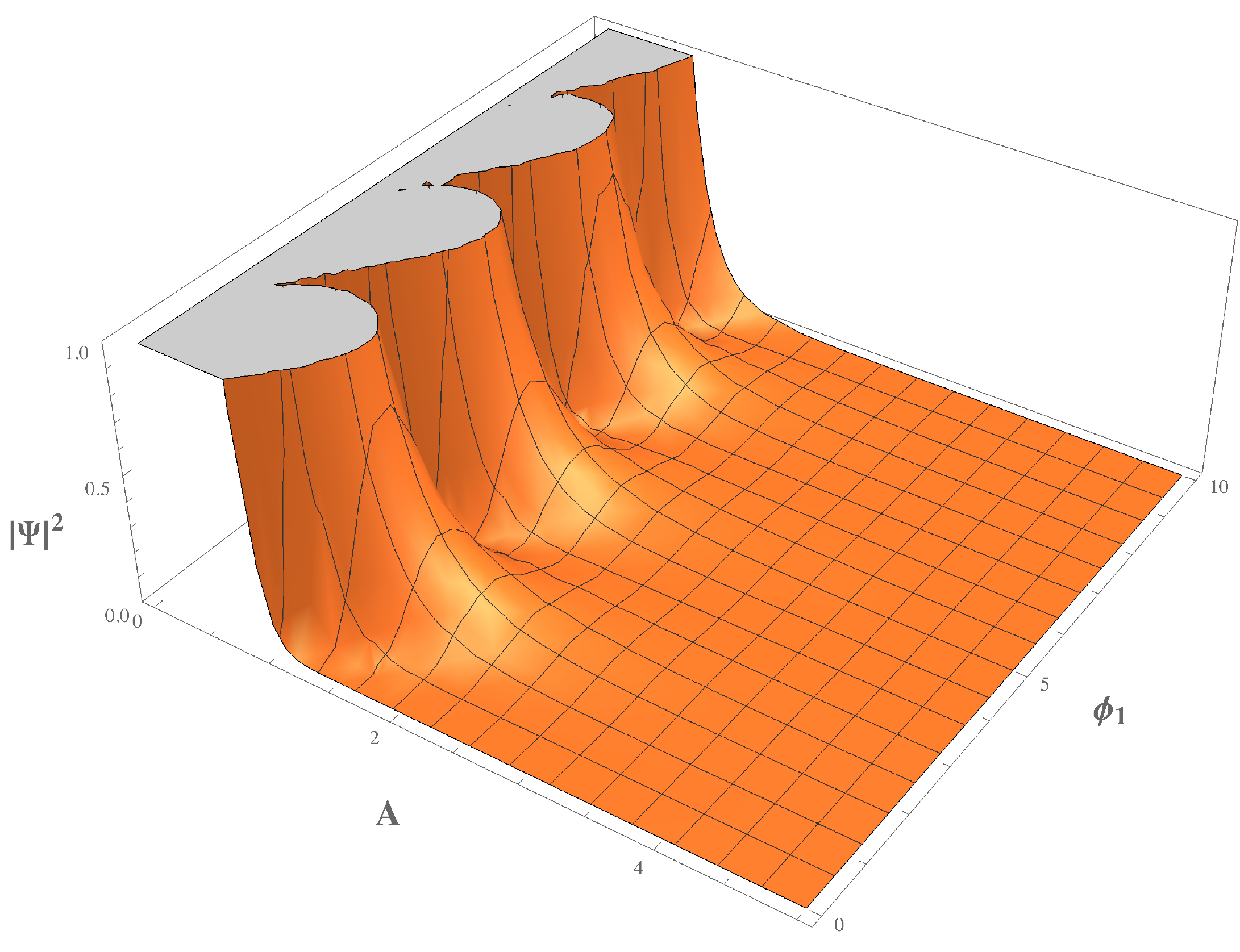}
\includegraphics[scale=0.35]{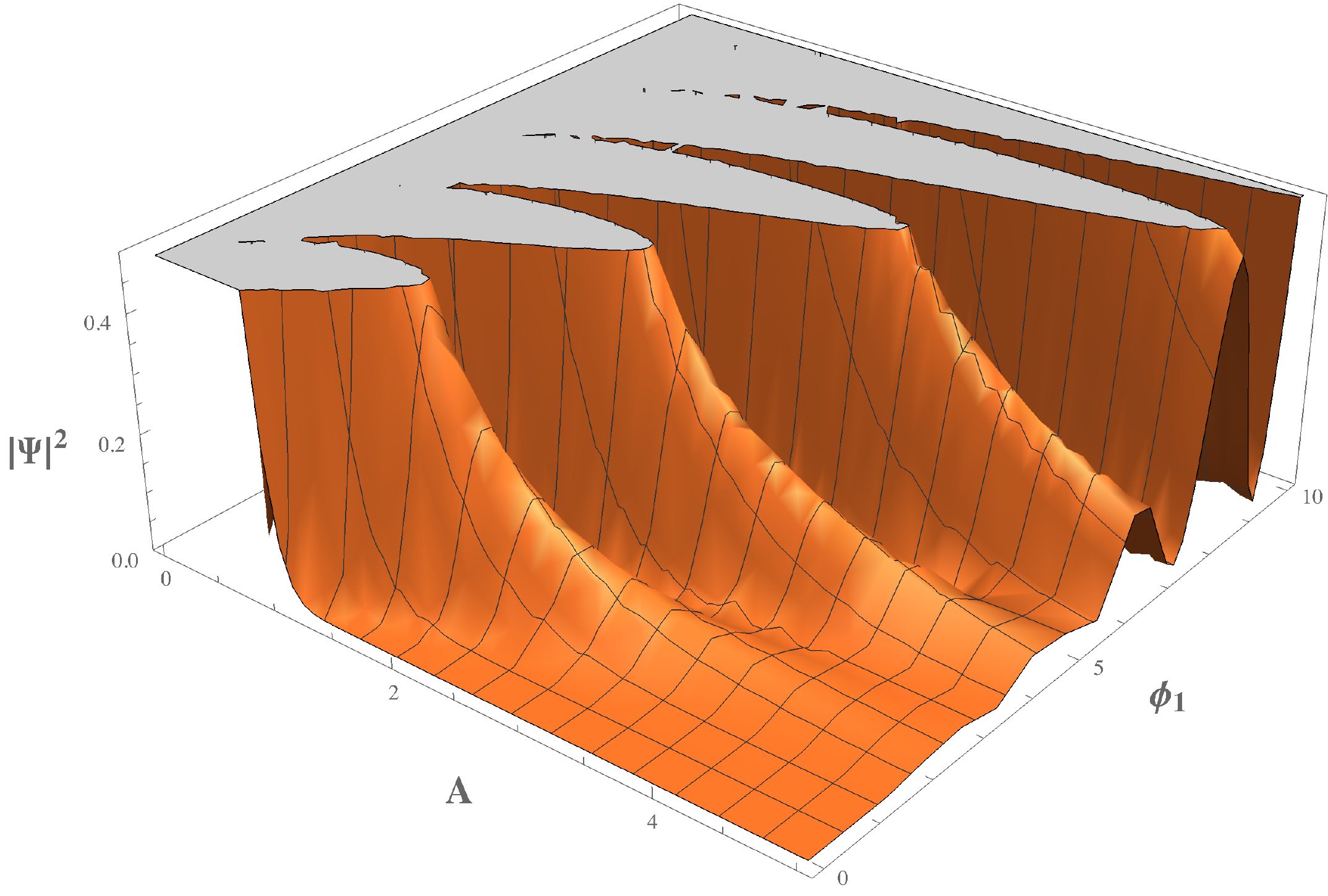}
\caption{Behaviour of density probability for $\lambda_1^2>\sqrt{3}$.
For both pictures: $\lambda_1=6$, $a_2=2$, $a_3=1$ where  $\rm Q=-2$ and $\rm Q=-10$ are the values for the top and bottom figures, respectively.}
\label{density-modifyQ}
\end{center}
\end{figure}
The behavior of the wave function for $\lambda_1^2>{3}$ can be seen
in Fig.~(\ref{density-modifyQ}), where the probability density
presents a damping behavior with respect to the scale factor, which
as mention before, is a desire characteristic in a wave function.
Here, the parameter $\rm Q$, for negative values, plays the role of a retarder of the wave
function and compresses the length on the axis
where the field evolves; then the inflation epoch should also be
retarded as time evolves.


\subsection{Quantum quintom like case}\label{quantum_quintom}
\begin{figure}[ht!]
\begin{center}
\captionsetup{width=.9\textwidth}
\includegraphics[scale=0.6]{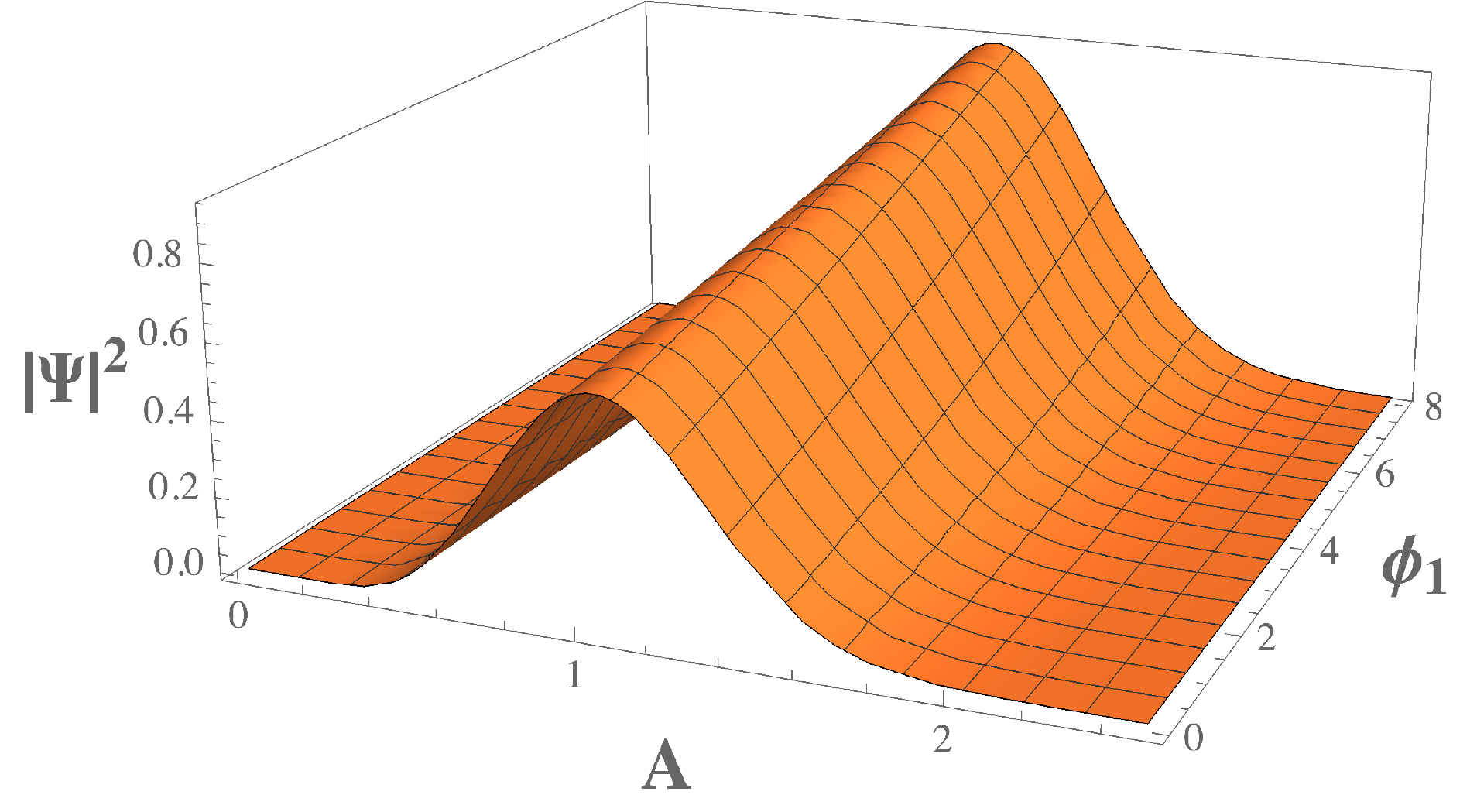}
\caption{Probability density for the quintom like cosmology; with $\ell=1$, $\rm a_3=1$, $\alpha=1$, $\rm \delta_{\ell}=1.207$, $\beta_{\ell}=1.414$, $\lambda_1=0.1$, $\mu_{\ell}=0.019$, $\rm c_{\ell}=0.08$.}\label{densidad_quintom}
\end{center}
\end{figure}

For the second cosmological model, the quintom like
case, the quantum version of this model is obtained applying, again, the recipe $\rm \Pi_{q^\mu}=-i\hbar \partial_{q^\mu}$ to
the Hamiltonian density (\ref{hamifrwa}), hence
\begin{equation}\label{qhamifrwa}
\begin{split}
\rm \biggl[ \frac{\hbar^2}{\mu_{_\ell}}\frac{\partial^2}{\partial
\xi_1^2} +\frac{\hbar^2}{\mu_{_\ell}}\frac{\partial^2}{\partial
\xi_2^2} -\hbar^2\left(48-\frac{1}{3c_{_\ell}}
\right)\left(\frac{\partial^2}{\partial \xi_3 \partial \xi_1} +
\frac{\partial^2}{\partial \xi_3 \partial \xi_2}
 \right)
-\hbar^2\left(16-\frac{1}{18c_{_\ell}}\right)\frac{\partial^2}{\partial
\xi_3^2}\\
-24V_1e^{-\xi_1}-24V_2 e^{-\xi_2}\biggr] \Psi=0,
\end{split}
\end{equation}
because the scalar potential does not depend on the coordinate $\rm
\xi_3$, we propose the following ansatz for the wave function $\rm
\Psi(\xi_1,\xi_2,\xi_3)=e^{(a_3/\hbar)\xi_3}{\cal A}(\xi_1) {\cal
B}(\xi_2)$ where $\rm a_3$ is an arbitrary constant. Substituting
and dividing by the ansatz in (\ref{qhamifrwa}), we obtain
\begin{equation}
\begin{split}
\rm \frac{\hbar^2}{\mu_{_\ell}{\cal A}} \frac{d^2 {\cal
A}}{d\xi_1^2}+ \frac{\hbar^2}{\mu_{_\ell}{\cal B}} \frac{d^2 {\cal
B}}{d\xi_2^2}-a_3 \hbar \left(48-\frac{1}{3c_{_\ell}}
\right)\left(\frac{1}{{\cal A}}\frac{d{\cal A}}{d \xi_1} +
\frac{1}{{\cal B}} \frac{d{\cal B}}{d \xi_2} \right)-a_3^2\left(16-\frac{1}{18c_{_\ell}}\right)\\
-24V_1e^{-\xi_1}-24V_2 e^{-\xi_2}=0,
\end{split}
\end{equation}
where we can separate the equations as
\begin{align}
\rm \frac{d^2 {\cal A}}{d\xi_1^2}- \frac{a_3 \mu_{_\ell}}{\hbar}
\left(48-\frac{1}{3c_{_\ell}} \right)\frac{d{\cal A}}{d \xi_1}
-\frac{\mu_{_\ell}}{\hbar^2} \left(\frac{a_3^2}{2}
\left(16-\frac{1}{18c_{_\ell}}\right)-\alpha^2+24V_1e^{-\xi_1}
\right) {\cal A}=0, \label{a}\\
\rm \frac{d^2 {\cal B}}{d\xi_2^2}-\frac{a_3 \mu_{_\ell} }{\hbar}
\left(48-\frac{1}{3c_{_\ell}} \right)\frac{d{\cal B}}{d \xi_2}
-\frac{\mu_{_\ell}}{\hbar^2}\left(\frac{a_3^2}{2}
\left(16-\frac{1}{18c_{_\ell}}\right)+\alpha^2+24V_2e^{-\xi_2}\right)
{\cal B}=0, \label{b}
\end{align}
with $\alpha^2$ being the separation constant. The corresponding solutions of (\ref{a}) and (\ref{b}) have the following form \cite{polyanin}
\begin{equation}
\rm Y(x)=Exp\left({-\frac{ax}{2}}\right)
Z_\nu\left(\frac{2\sqrt{b}}{\lambda} e^{\frac{\lambda x}{2}}\right),
\end{equation}
here $\rm Z_\nu$ are the generic Bessel function with order $\rm \nu=\sqrt{a^2-4c}/\lambda$. If $\sqrt{b}$ is
real, $\rm Z_\nu$ are the ordinary Bessel function, otherwise the solution will be given by the modified Bessel function.
Making the following identifications
\begin{eqnarray}
\rm \lambda&=&-1, \\
\rm a&=&-\frac{a_3 \mu_{_\ell}}{\hbar}\left(48-\frac{1}{3c_{_\ell}} \right), \\
\rm b_{1,2}&=&-\frac{\mu_{_\ell}}{\hbar^2}24V_{1,2}, \\
\rm c_\mp&=&-\frac{\mu_{_\ell}}{\hbar^2}\left(a_3^2\left(8-\frac{1}{36c_{_\ell}}\right)\mp \alpha^2\right),\\
\rm \nu_\mp&=&\sqrt{\frac{\rm a^2}{\mu_\ell}+4\rm c_\mp},
\end{eqnarray}
we can check that the value for $\sqrt{b}$ is imaginary, which as
already mentioned, gives a solution in terms of the modified Bessel
function $\rm Z_\nu=K_\nu$ whose order lies in the reals. Thus, the
wave function is
\begin{equation}\label{qsol_quintom}
\begin{split}
\rm \Psi_{\nu_\pm}=
Exp\left[\left(\frac{\mu_{_\ell}}{2\hbar}\left(48-\frac{1}{3c_{_\ell}}\right)(\xi_1+\xi_2)+\frac{\xi_3}{\hbar}\right)a_3\right]
\rm K_{\nu_-}\left(\frac{4}{\hbar}\sqrt{6V_1\mu_{_\ell}}
e^{-\frac{\xi_1}{2}} \right)\times\\
K_{\nu_+}\left(\frac{4}{\hbar}\sqrt{6V_2\mu_{_\ell}}
e^{-\frac{\xi_2}{2}} \right).
\end{split}
\end{equation}
so, the wave function in the original variables becomes
\begin{equation}\label{original_quintom}
\begin{split}
\rm \Psi_{\nu_\pm}= A^{-\frac{4a_3\delta_\ell}{\hbar}}\,
Exp\left\{\frac{\beta_\ell a_3}{6\hbar}\left(\lambda_1 \phi_1
+\lambda_12 \phi_2 \right) \right\} \rm
K_{\nu_-}\left(\frac{4}{\hbar}\sqrt{6V_1\mu_{_\ell}}
A^3e^{-\frac{\lambda_1}{2}\phi_1} \right)\times\\
\rm K_{\nu_+}\left(\frac{4}{\hbar}\sqrt{6V_2\mu_{_\ell}}
A^3e^{-\frac{\lambda_2}{2}\phi_2} \right),
\end{split}
\end{equation}
where $\rm
\delta_\ell=\frac{3}{2}\mu_\ell\left(48-\frac{1}{3c_\ell}\right)+1$
and $\rm \beta_\ell=3\mu_\ell\left(48-\frac{1}{3c_\ell} \right)+1$.
In Fig.~(\ref{densidad_quintom}) the behavior of the probability
density with respect of the scale factor and the scalar field
$\phi_1$ is depicted; we can note that as the scale factor evolves
the probability density has a decaying behavior and and has a
moderate growth in the axis where the scalar field develops (a
similar behavior was found for the quantum solution $\lambda_1^2>3$
when $\rm Q=-10$).
\\

\subsection{Quantum quintessence like case}
\begin{figure}[ht!]
\begin{center}
\captionsetup{width=.9\textwidth}
\includegraphics[scale=0.6]{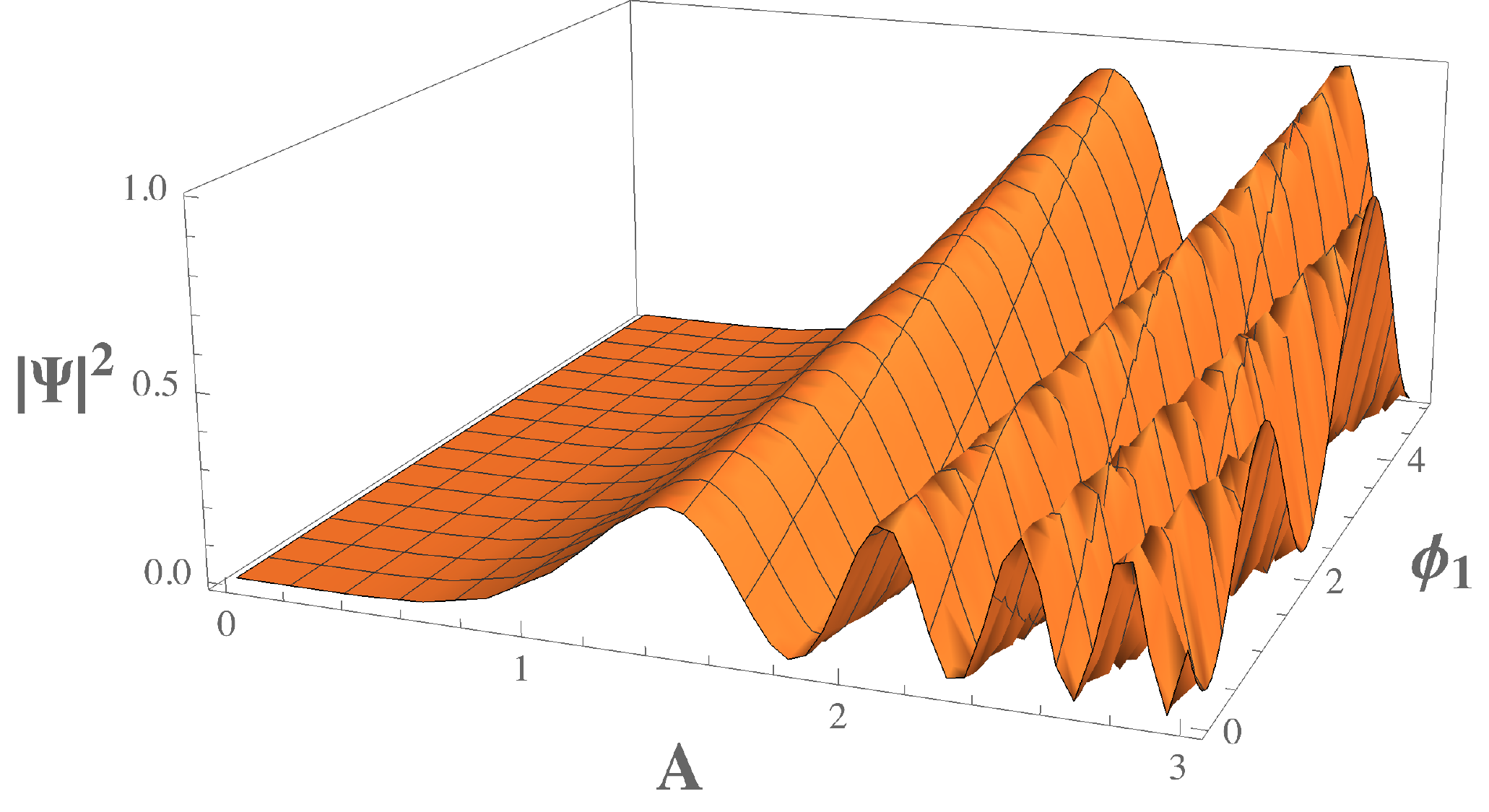}
\caption{Probability density for the quintessence like cosmological model, where $\ell=1$, $\rm a_3=-1$, $\alpha=0.1$, $\rm \delta_{\ell}^{\prime}=-0.792$, $\beta_{\ell}^{\prime}=0.585$, $\lambda_1=0.3$, $\mu_{\ell}^{\prime}=0.019$, $\rm c_{\ell}^{\prime}=0.08$.}\label{densidad_quintessence}
\end{center}
\end{figure}

Lastly, we are going to consider the quantum version of the quintessence like case. As in the previous two models,
what we want is to obtain an equation of the form $\rm {\cal H}\Psi(\xi_i)=0$, to achieve this we introduce the standard prescription
$\Pi_q^\mu=-i\hbar\partial_{q^\mu}$ in (\ref{hamifrwb}), obtaining
\begin{equation}
\begin{split}
\rm \biggl[ -\frac{\hbar^2}{\nu_{_\ell}}\frac{\partial^2}{\partial
\xi_1^2} -\frac{\hbar^2}{\nu_{_\ell}}\frac{\partial^2}{\partial
\xi_2^2} -\hbar^2\left(48-\frac{1}{3c_{\ell}^{\prime}}
\right)\left(\frac{\partial^2}{\partial \xi_3 \partial \xi_1} +
\frac{\partial^2}{\partial \xi_3 \partial
\xi_2}\right)-\hbar^2\left(16-\frac{1}{18c_{\ell}^{\prime}}\right)\frac{\partial^2}{\partial
\xi_3^2}\\
-24V_1e^{-\xi_1}-24V_2 e^{-\xi_2}\biggr] \Psi=0,
\label{qhamifrwb}
\end{split}
\end{equation}
we can see that the scalar potential does not depend on the coordinate $\rm
\xi_3$, consequently we propose the following ansatz for the wave function $\rm
\Psi(\xi_1,\xi_2,\xi_3)=e^{(b_3/\hbar)\xi_3}{\cal A}(\xi_1)
{\cal B}(\xi_2)$ where $\rm b_3$ is an arbitrary constant.
Applying and dividing by the ansatz in (\ref{qhamifrwb}) we get
\begin{equation}
\begin{split}
\rm -\frac{\hbar^2}{\nu_{_\ell}{\cal A}} \frac{d^2 {\cal
A}}{d\xi_1^2}- \frac{\hbar^2}{\nu_{_\ell}{\cal B}} \frac{d^2 {\cal
B}}{d\xi_2^2}-b_3 \hbar \left(48-\frac{1}{3c_{\ell}^{\prime}}
\right)\left(\frac{1}{{\cal A}}\frac{d{\cal A}}{d \xi_1} +
\frac{1}{{\cal B}} \frac{d{\cal B}}{d \xi_2} \right) -b_3^2
\left(16-\frac{1}{18c_{\ell}^{\prime}}\right)\\
-24V_1e^{-\xi_1}-24V_2 e^{-\xi_2}=0,
\end{split}
\end{equation}
separating the equations we have that
\begin{align}
\rm \frac{d^2 {\cal A}}{d\xi_1^2}+ \frac{b_3 \nu_{_\ell}}{\hbar}
\left(48-\frac{1}{3c_{\ell}^{\prime}} \right)\frac{d{\cal A}}{d \xi_1}
+\frac{\nu_{_\ell}}{\hbar^2} \left(b_3^2
\left(8-\frac{1}{36c_{_\ell}^{\prime}}\right)-\alpha^2+24V_1e^{-\xi_1}
\right) {\cal A}=0, \label{ba}\\
\rm \frac{d^2 {\cal B}}{d\xi_2^2}+\frac{b_3 \nu_{_\ell} }{\hbar}
\left(48-\frac{1}{3c_{_\ell}^{\prime}} \right)\frac{d{\cal B}}{d \xi_2}
+\frac{\nu_{_\ell}}{\hbar^2}\left(b_3^2
\left(8-\frac{1}{36c_{_\ell}^{\prime}}\right)+\alpha^2+24V_2e^{-\xi_2}\right)
{\cal B}=0, \label{bb}
\end{align}
where $\alpha^2$ is the separation constant. These last two equations are similar to the quantum quintom like case
(\ref{a}) and (\ref{b}). Proceeding in a similar fashion as the previous subsection \ref{quantum_quintom}, we make the following identifications
\begin{eqnarray}
\rm \lambda&=&-1, \\
\rm a&=&\frac{b_3 \nu_{_\ell}}{\hbar}\left(48-\frac{1}{3c_{_\ell}^{\prime}} \right), \\
\rm b_{1,2}&=&\frac{\nu_{_\ell}}{\hbar^2}24V_{1,2}, \\
\rm c_\mp&=&\frac{\nu_{_\ell}}{\hbar^2}\left(b_3^2\left(8-\frac{1}{36c_{_\ell}^{\prime}}\right)\mp \alpha^2\right), \\
\end{eqnarray}
and conclude that the solutions are given by the ordinary Bessel
function $J_\nu$ with order $\rm\nu_\mp=\sqrt{(\rm
a^2/\nu_\ell)+4\rm c_\mp}$.
Thus, the wave function becomes
\begin{equation}\label{qsol_quintessence}
\begin{split}
\rm \Psi_{\nu_\pm}=
Exp\left[\left(\frac{\nu_{_\ell}}{2\hbar}\left(48-\frac{1}{3c_{_\ell}^{\prime}}\right)(-\xi_1-\xi_2)+\frac{\xi_3}{\hbar}\right)b_3\right]
\rm J_{\nu_-}\left(\frac{4}{\hbar}\sqrt{6V_1\nu_{_\ell}}
e^{-\frac{\xi_1}{2}} \right)\times\\
J_{\nu_+}\left(\frac{4}{\hbar}\sqrt{6V_2\nu_{_\ell}}
e^{-\frac{\xi_2}{2}} \right).
\end{split}
\end{equation}
written in the original variables, become
\begin{equation}\label{original_quintessence}
\begin{split}
\rm \Psi_{\nu_\pm}= A^{\frac{4b_3\delta_\ell}{\hbar}}\,
Exp\left\{-\frac{\beta_\ell b_3}{6\hbar}\left(\lambda_1 \phi_1
+\lambda_12 \phi_2 \right) \right\} \rm
J_{\nu_-}\left(\frac{4}{\hbar}\sqrt{6V_1\nu_{_\ell}}
A^3e^{-\frac{\lambda_1}{2}\phi_1} \right)\times\\
\rm J_{\nu_+}\left(\frac{4}{\hbar}\sqrt{6V_2\nu_{_\ell}}
A^3e^{-\frac{\lambda_2}{2}\phi_2} \right).
\end{split}
\end{equation}
where $\rm
\delta_\ell=\frac{3}{2}\nu_{_\ell}\left(48-\frac{1}{3c_{\ell}^{\prime}}\right)-1$
and $\rm\beta_\ell= -3\nu_{_\ell}
\left(48-\frac{1}{3c_{\ell}^{\prime}} \right)+1$.
Fig.~(\ref{densidad_quintessence}) shows the probability density for
the quintessence cosmological model in terms of the scale factor and
the scalar field $\phi_1$; we can see that the probability density
dies away in a muffled manner with respect of the scale factor,
which is a good characteristic of a wave function, as pointed out in
section (\ref{quantum_QK}). Another thing we can notice is that the
probability density has a moderate increase in the direction were
the scalar field evolves (as in the quantum quintom case).
\section{Final Remarks}\label{conclusions}
In this work we have studied two multi-field cosmological models,
for which classical and quantum solutions were found. For both
setups we work with a flat FRW cosmology. In the first model, we
consider two scalar fields but only one potential term and standard
matter in the stiff scenario, which can be seen as a simple
quintessence plus a K-essence model. In the second one, we also
considered two scalar fields with the difference that the two
potential terms are taken into account, as well as the standard
kinetic energy and the mixed term, which are present in chiral field
approach and when standard matter is included it can be thought of
as a stiff matter scenario. Regarding this second model, it is shown
that two possible cases can be studied: a quintom like model and a
quintessence like model considering the mixed term in the scalar
fields.

For the first flat FRW model, applying the Hamiltonian approach, we
where able to find three different classical solutions depending on
the value of the parameter $\lambda_1^2$. In each of the three
cases, to analyze the dynamical behavior of the model under
consideration,
 we calculate the deceleration parameter $\rm q_i$ (where $\rm q_1, q_2$ and $\rm q_3$ stand for the solutions for
$\lambda_1^2<3$, $\lambda_1^2>3$ and $\lambda_1^2=3$, respectively)
when no matter is present and when standard matter is included. In
Fig.~(\ref{q-parameters}) we can observe the temporal evolution of
the deceleration parameters, in the case for $\rho_\gamma=0$ it is
observed that $\rm q_1$ and $\rm q_3$ have a
deceleration/acceleration phase, where for both parameters the
former is for a short period of time to then accelerate, whereas in
the case when standard matter is included ($\rho_\gamma\not=0$), the
behavior of deceleration/acceleration persist, but $\rm q_3$ gets
outgrown by $\rm q_1$. For the case of $\rm q_2$, in contrast to
$\rm q_1$ and $\rm q_3$, an accelerated behavior is only present and
a slow growth in both scenarios with no significant difference is
shown. Ultimately, in both scenarios, the asymptotic behavior of the
three solutions goes to the same value, $-1$. In
Fig.~(\ref{omegas-polvo}) we can observe the behavior of the
barotropic parameter (in the gauge $\rm N\not=1$) for the three
different solutions of $\lambda_1^2$, for each of the three
parameters the asymptotic behavior approaches $1$. For completeness,
we also calculate the barotropic parameter for the gauge $\rm N=1$,
given by Eq.~(\ref{norma}). In this model, multiplying the kinetic
energy term associated to the second scalar field by an arbitrary
function $F(\phi_2)$, would be an interesting exercise to see if
exact solutions can be found; research in this direction has been
performed in \cite{soc-abraham1,soc-abraham2}. Another avenue that
can be explored is the one presented in \cite{Sa}, as well the ideas
pursued in \cite{Dimakis:2019qfs, Paliathanasis:2014yfa}. In section
$\ref{2.2}$ we investigate the case when standard matter is
included, in this particular case we found the scale factor of the
Universe (\ref{solution-a}) has an accelerated growth and whose
dynamical demeanor can be observed in Fig.~(\ref{volumen_polvo}),
this characteristic is also corroborated with the deceleration
parameter given by (\ref{q-gama-difcero}). We also found that the
scalar potential, given by (\ref{potential_phi}), is consistent with
a (volume) accelerated expansion in the dust scenario. In addition,
we found that the scalar field $\phi_2$ acquires a constant value
for late times, this feature was also obtained for an anisotropic
cosmological model, where the anisotropic parameters vanished for
late times \cite{los4}. To round off this analysis we calculate the
barotropic parameter, given by (\ref{omega-difcero}). Finally, we
study the case for $\gamma=\frac{1}{3}$, which corresponds to the
radiation era. The solution of the master equation for this case is
given by (\ref{solucion_radiacion}), unfortunately the expression
for the scale factor is not given explicitly in terms of $\rm t$,
given that the solution obtained is not invertible, nevertheless,
one would expect (in light of (\ref{A_hiper})) that if we could get
$\rm A(t)$ it would be a function with an accelerated growth.

The quantum version of these model
 were obtained making the usual replacement $\rm \Pi_{q^\mu}=-i\hbar \partial_{q^\mu}$ in the classical
  Hamiltonian density, where the linear term
  $\rm  e^{-3\Omega}\Pi_\Omega^2 \to e^{-3\Omega}\left[\Pi_\Omega^2 +Qi\hbar \Pi_\Omega \right]$ was introduced,
  in order to account for the factor ordering problem, where $\rm Q$ is a real number that measures the ambiguity
  in the factor ordering. In this set up we found that, for $\lambda_1^2<3$, the wave function has a damping behavior,
  which is a good characteristic that has also been reported in \cite{soco2, sor}, features that can be seen
  in Fig.~(\ref{density1}) and Fig.~(\ref{cortes}). The quantum solution for $\lambda_1^2>3$ is given Eq.~(\ref{k1}),
   and the behavior of the wave function is presented in Fig.~(\ref{density-modifyQ}), for which the
   damping behavior remains, with the difference that for negative values, the parameter $\rm Q$ plays
   the role of a retarder of the wave function and the length of the scalar field is compressed, signaling that
   the inflation period should also be retarded over time.

For the second model, in addition to considering the scalar fields,
the two terms of the potential were also considered. First, we were
able to find classical solutions to the EKG equations
(\ref{mono}-\ref{ekg-phi}) using the Hamiltonian formalism. In this
model we were able to distinguish two types of solutions: a quintom
type and a quintessence type.  For the first, the solutions are
given by the equations (\ref{sols_quintom}) while for the second are
given by (\ref{sols_quintessence}), with these two sets of solutions
the scale factor, deceleration parameter and the barotropic
parameter could be found. In Fig.~(\ref{vol_q_omega_ambos}) we can
see the behavior of the volume function, the q-parameters and the
barotropic parameters for the models. The volume function for both
models has an accelerated growth. The deceleration parameter for the
quintom case increases more rapidly than quintessence counterpart to
then stabilizing at $-1$; finally the barotropic parameter for the
quintom model acquires faster the asymptotic value of $1$. Quantum
solutions for
 to this model were also found. The solution for the
quantum quintom like model is given by (\ref{original_quintom}); in
Fig.~(\ref{densidad_quintom}) can be appreciated that the
probability density drops as the scale factor develops while in the
direction of the scalar field it has a steady increment. The
solution for the quantum quintessence like case, is given by
(\ref{original_quintessence}) and the probability density its shown
in Fig.(\ref{densidad_quintessence}), where the probability density
dies away in a damped manner as the scale factor evolves, also a
slight increase in the the direction of the scalar function is
observed. Finally, we can say that this work has already been done
considering the anisotropic bianchi type I and the solutions found
are generalization of the solutions to this work \cite{SSAL}.

 In the
page below, three tables are presented where our results are
included.

\begin{table}
\caption{Table 1: First Model (classical)}
    \centering
\begin{turn}{90}
{\tiny
\begin{tabular}{|c|c|c|c|}
     \hline
     Cases& Scale Factor & $\rm q-parameter$ & $\omega_T$ \\  [0.5ex]
     \hline
     $\lambda_1^2<3$ & $\rm A(t)=A_{0}\exp\left[-\frac{12\lambda_1p_{\phi_1}}{\eta}(t-t_0)\right]\left(Csch\left[12\omega(t-t_0)\right]\right)^{\frac{1}{\eta}}$ &  $\rm q_1=\rm -1-\frac{\eta\omega^2}{\left[\lambda_1p_{\phi_1}Sinh(12\omega t)+\omega Cosh(12\omega t)\right]^2}$ & $\rm\omega_{T_1}=\rm 1-\frac{2\eta \omega^2}{3\left[\lambda_1p_{\phi_1}Sinh(12\omega t)+\omega Cosh(12\omega t)\right]^2}$ \\ [0.5ex]
     \hline
     $\lambda_1^2>3$ & $\rm A(t)=A_0\, Exp\left[ \frac{12\lambda_1 p_{\phi_1}}{\beta}(t-t_0)
\right] \,\,Cosh^{\frac{1}{\beta}}\left(12\omega_2(t-t_0)\right)$ &
$\rm q_2=\rm
-1-\frac{\beta\omega^{\prime2}}{\left[\lambda_1^{\prime}p_{\phi_1}Cosh(12\omega^{\prime}
t)+\omega^{\prime} Sinh(12\omega^{\prime} t)\right]^2}$ & $\rm
\omega_{T_2}=\rm 1-\frac{2}{3}\frac{\beta
{\omega^\prime}^2}{\left[\lambda_1^{\prime}p_{\phi_1}Cosh(12\omega^{\prime}
t)+\omega^{\prime} Sinh(12\omega^{\prime} t)\right]^2}$  \\ [0.5ex]
     \hline
     $\lambda_1^2=3$ & $\rm A(t)=A_0 Exp\left[2\sqrt{3}\frac{p_{\phi_1}^2+p_{\phi_2}^2+4\rho_1}{p_{\phi_1}}(t-t_0)\right]Exp\left[ \frac{\sqrt{3}p}{36p_{\phi_1}}e^{24\sqrt{3}p_{\phi_1}(t-t_0)}\right]$ & $\rm q_3=\rm -1-\frac{12\sqrt{3}pp_{\phi_1}^3
Exp(24\sqrt{3}p_{\phi_1}t)}{[pp_{\phi_1}Exp(24\sqrt{3}p_{\phi_1}t)+\sqrt{3}(p_{\phi_1}^2+p_{\phi_2}^2+4\rho_1)]^2}$
& $\rm \omega_{T_3}=\rm 1-\frac{2}{3}\frac{12\sqrt{3}pp_{\phi_1}^3
Exp(24\sqrt{3}p_{\phi_1}t)}{[pp_{\phi_1}Exp(24\sqrt{3}p_{\phi_1}t)+\sqrt{3}(p_{\phi_1}^2+p_{\phi_2}^2+4\rho_1)]^2}$\\
[0.5ex]
     \hline
 $\gamma=0$ & $\rm A^3(t)=a_0\left[(a_1 t +a_2)^2-1\right]$ &$\rm q(t)=\rm -\frac{1}{2} +\frac{1}{2(a_1 t +a_2)^2}$ & $\rm\omega_T=\frac{2}{3}q-\frac{1}{3}=\frac{1}{3(a_1 t +a_2)^2}$\\ [0.5ex]
     \hline
 $\gamma=\frac{1}{3}$ & $\rm 2b_0\,\Delta t=A\sqrt{b_0\,A^2+b_2}-\frac{b_2}{\sqrt{b_0}}\, Ln\left[A+\sqrt{A^2+\frac{b_2}{b_0}}\right]$ & & \\ [0.5ex]
     \hline
 \end{tabular}
 }
 \end{turn}
\end{table}

\begin{table}
\caption{Table 2: Second Model (classical)}
    \centering
\begin{turn}{90}
{\tiny
 \begin{tabular}{|c|c|c|c|}
     \hline
     Cases& Scale Factor & $\rm q-parameter$ & $\omega_T$ \\  [0.2ex]
     \hline
     $quintom$ & $\rm A(t)=A_0\, Cosh^{\frac{1}{6}+\alpha_q}\left(r_1\,t-q_1\right)\,Cosh^{\frac{1}{6}+\alpha_q}\left(r_2\,t-q_2 \right)e^{-\beta_q\,\Delta t}$ &  $\rm q_{quintom}=-1 -\alpha_0\,\frac{\left(r_1^2\,Cosh^2(r_2\,t-q_2)+r_2^2\,Cosh^2(r_1\,t -q_1)\right)}{T}$ & $\rm \omega_{quintom}=1 -\frac{2}{3}\alpha_0\,\frac{\left(r_1^2\,Cosh^2(r_2\,t
-q_2)+r_2^2\,Cosh^2(r_1\,t -q_1)\right)}{T}$ \\ [0.2ex]
     \hline
     $quintessence$ & $\rm A(t)=A_0\, Sinh^{\frac{1}{6}-\alpha_q}\left(r_1\,t-q_1\right)\,Sinh^{\frac{1}{6}-\alpha_q}\left(r_2\,t-q_2 \right)e^{\beta_q\,\Delta t}$ & $\rm q_{quintessence}=-1 -\frac{\alpha_1\left(r_1^2\,Sinh^2(r_2\,t
-q_2)+r_2^2\,Sinh^2(r_1\,t -q_1)\right)}{Q}$ & $\rm
\omega_{quintessence}=1-\frac{2}{3}\frac{\alpha_1\left(r_1^2\,Sinh^2(r_2\,t-q_2)+r_2^2\,Sinh^2(r_1\,t
-q_1)\right)}{Q}$  \\ [0.2ex]
     \hline
     \end{tabular}
}
 \end{turn}
\end{table}

\begin{table}
\caption{Table 3: Quantum solutions (first two lines correspond to
the first model and the second two lines to the second model)}
    \centering
\begin{turn}{90}
{\tiny
     \begin{tabular}{|c|c|}
     \hline
        Cases & Wave Function  \\ [0.5ex]
     \hline
     $\lambda_1^2<3$ & $\rm \Psi_{\nu_2} =\rm  A^{-\alpha_2 }\, Exp\left(-\frac{2a_2+\hbar Q}{4\hbar(\lambda_1^2-3)}\lambda_1 \phi_1 - \frac{a_3}{\hbar}\phi_2 \right)J_{\nu_2}\left(\frac{2}{\hbar}\sqrt{\frac{2V_0}{ 3-\lambda^2}}A^{3}e^{\frac{\lambda_1}{2}\phi_1} \right)$  \\ [0.5ex]
     \hline
     $\lambda_1^2>3$ & $\rm \Psi_{\nu_1} =\rm  A^{-\alpha_1 }Exp\left(\frac{2a_2+ \hbar Q}{4\hbar(\lambda_1^2-3)}\lambda_1 \phi_1 - \frac{ a_3}{\hbar}\phi_2\right) K_{\nu_1}\left(
\frac{2}{\hbar}\sqrt{\frac{2V_0}{\lambda_1^2-3}}A^{3}e^{\frac{\lambda_1}{2}\phi_1}
\right)$  \\ [0.5ex]
     \hline
     $quintom$ & $\rm \Psi_{\nu_\pm}= A^{-\frac{4a_3\delta_\ell}{\hbar}}Exp\left\{\frac{\beta_\ell a_3}{6\hbar}\left(\lambda_1 \phi_1+\lambda_12 \phi_2 \right) \right\} \rm K_{\nu_-}\left(\frac{4}{\hbar}\sqrt{6V_1\mu_{_\ell}}A^3e^{-\frac{\lambda_1}{2}\phi_1} \right) \rm K_{\nu_+}\left(\frac{4}{\hbar}\sqrt{6V_2\mu_{_\ell}}A^3e^{-\frac{\lambda_2}{2}\phi_2} \right)$ \\ [0.5 ex]
     \hline
     $quintessence$ & $\rm \Psi_{\nu_\pm}= A^{\frac{4b_3\delta_\ell}{\hbar}}Exp\left\{-\frac{\beta_\ell b_3}{6\hbar}\left(\lambda_1 \phi_1+\lambda_12 \phi_2 \right) \right\} \rm J_{\nu_-}\left(\frac{4}{\hbar}\sqrt{6V_1\nu_{_\ell}}A^3e^{-\frac{\lambda_1}{2}\phi_1} \right)\rm J_{\nu_+}\left(\frac{4}{\hbar}\sqrt{6V_2\nu_{_\ell}}A^3e^{-\frac{\lambda_2}{2}\phi_2} \right)$ \\ [0.5ex]
     \hline
    \end{tabular}
 }
 \end{turn}
\end{table}

\acknowledgments{  \noindent This work was partially supported by
PROMEP grants UGTO-CA-3. J.S. was partially supported SNI-CONACYT.
This work is part of the collaboration within the Instituto Avanzado
de Cosmolog\'{\i}a and Red PROMEP: Gravitation and Mathematical
Physics under project {\it Quantum aspects of gravity in
cosmological models, phenomenology and geometry of space-time}. Many
calculations where done by Symbolic Program REDUCE 3.8. We also want
to thank the anonymous referees for their valuable recommendations.}


\end{document}